\documentclass[twocolumn]{aastex631}
\usepackage{amsbsy}
\usepackage{amsfonts}
\usepackage{amssymb}
\usepackage{bm,ulem}
\usepackage{mathrsfs}
\usepackage{pifont}
\usepackage{stmaryrd}
\usepackage{textcomp,bookmark}
\usepackage{xspace}
\usepackage{amsmath,amsxtra}
\usepackage[OT2,OT1]{fontenc}
\usepackage{graphicx}
\usepackage{hyperref}
\usepackage{natbib}
\usepackage{array,colortbl,diagbox}
\usepackage{harpoon}
\usepackage{ulem}
\usepackage{xcolor}

\newcommand{\msun}{\ifmmode {{M}_\odot} \else  {${\rm M}_\odot$} \fi}
\newcommand{\hms}{\ifmmode {h^{-1}{\rm M}_\odot} \else  {$h^{-1}{\rm M}_\odot$} \fi}
\newcommand{\mpc}{\ifmmode {{\rm Mpc}} \else  {${\rm Mpc}$} \fi}
\newcommand{\kpc}{\ifmmode {{\rm kpc}} \else  {${\rm kpc}$} \fi}
\newcommand{\hmpc}{\ifmmode {h^{-1}{\rm Mpc}} \else  {$h^{-1}{\rm Mpc}$} \fi}
\newcommand{\ksm}{\ifmmode {{\rm km}~{\rm s}^{-1}~{\rm Mpc}^{-1}} \else  {${\rm km}~{\rm s}^{-1}~{\rm Mpc}^{-1}$} \fi}
\newcommand{\hkpc}{\ifmmode {h^{-1}{\rm kpc}} \else  {$h^{-1}{\rm kpc}$} \fi}
\newcommand{\amin}{\ifmmode {{\rm arcmin}} \else  {${\rm arcmin}$} \fi}
\newcommand{\mv}{\ifmmode {{M}_{\rm vir}} \else  {${M}_{\rm vir}$} \fi}
\newcommand{\mt}{\ifmmode {{M}_{200}} \else  {${M}_{200}$} \fi}
\newcommand{\mlow}{\ifmmode {{M}_{\rm low}} \else  {${M}_{\rm low}$} \fi}
\newcommand{\neff}{\ifmmode {{n}_{\rm eff}} \else  {${n}_{\rm eff}$} \fi}
\newcommand{\cv}{\ifmmode {{c}_{\rm vir}} \else  {${c}_{\rm vir}$} \fi}
\newcommand{\ct}{\ifmmode {{c}_{200}} \else  {${c}_{200}$} \fi}
\newcommand{\rv}{\ifmmode {{r}_{\rm vir}} \else  {${r}_{\rm vir}$} \fi}
\newcommand{\rt}{\ifmmode {{r}_{200}} \else  {${r}_{200}$} \fi}

\newcommand{\rdlt}{\ifmmode {{r}_{\scriptscriptstyle\Delta}} \else  {${r}_{\scriptscriptstyle\Delta}$} \fi}
\newcommand{\mdlt}{\ifmmode {{M}_{\scriptscriptstyle\Delta}} \else  {${M}_{\scriptscriptstyle\Delta}$} \fi}
\newcommand{\cdlt}{\ifmmode {{c}_{\scriptscriptstyle\Delta}} \else  {${c}_{\scriptscriptstyle\Delta}$} \fi}
\newcommand{\smM}{\ifmmode {\scriptscriptstyle M} \else  {${\scriptscriptstyle M}$} \fi}
\newcommand{\smT}{\ifmmode {\scriptscriptstyle T} \else  {${\scriptscriptstyle T}$} \fi}
\newcommand{\zbpz}{\ifmmode {z_{\scriptscriptstyle BPZ}} \else  {${z_{\scriptscriptstyle BPZ}}$} \fi}
\newcommand{\hh}{\ifmmode {h^{-1}} \else  {$h^{-1}$} \fi}
\newcommand{\vth}{\ifmmode {\vec{\theta}} \else  {$\vec{\theta}$} \fi}
\def\etal{et al.}
\def\ie{{\frenchspacing\it i.e.}}
\def\eg{{\frenchspacing\it e.g.}}

\newcommand{\remove}[1]{}
\newcommand{\yppn}{\ifmmode {{\gamma}_{\scriptscriptstyle {\rm PPN}}} \else  {${\gamma}_{\scriptscriptstyle {\rm PPN}}$} \fi}
\newcommand{\rpr}{\ifmmode {r^\prime} \else  {$r^\prime$} \fi}

\newcommand{\lgg}{\ifmmode {{\rm log}} \else  {${\rm log}$} \fi}
\newcommand{\sn}{\ifmmode {\sigma_n} \else  {$\sigma_n$} \fi}
\newcommand{\snsq}{\ifmmode {\sigma_n^2} \else  {$\sigma_n^2$} \fi}

\newcommand{\dd}{{\rm d}}

\begin{document}
\title{Mass Reconstruction of Galaxy-scale Strong Gravitational Lenses Using a Broken Power-law Model}

\author[0000-0001-9781-6863]{Wei Du}\thanks{E-mail: \url{duwei@shnu.edu.cn}}
\affiliation{Shanghai Key Lab for Astrophysics, Shanghai Normal University, Shanghai, 200234, China}

\author{Liping Fu}\thanks{E-mail: \url{fuliping@shnu.edu.cn}}
\affiliation{Shanghai Key Lab for Astrophysics, Shanghai Normal University, Shanghai, 200234, China}

\author[0000-0002-9063-698X]{Yiping Shu}
\affiliation{Purple Mountain Observatory, Chinese Academy of Science, Nanjing, 210023, China}

\author{Ran Li}
\affiliation{National Astronomical Observatories, Chinese Academy of Science, Beijing, 100101, China}
\affiliation{Institute for Frontiers in Astronomy and Astrophysics, Beijing Normal University,  Beijing, 102206, China}
\affiliation{School of Astronomy and Space Science, University of Chinese Academy of Sciences, Beijing, 100049, China}

\author[0000-0002-8397-012X]{Zuhui Fan}
\affiliation{South-Western Institute for Astronomy Research, Yunnan University, Kunming, 650500, China}

\author{Chenggang Shu}
\affiliation{Shanghai Key Lab for Astrophysics, Shanghai Normal University, Shanghai, 200234, China}

\begin{abstract}
  With mock strong gravitational lensing images, we investigate the performance of the broken power-law (BPL) model proposed by \citet{2020ApJ...892...62D} on the mass reconstruction of galaxy-scale lenses. An end-to-end test is carried out, including the creation of mock strong lensing images, the subtraction of lens light, and the reconstruction of lensed images, where the lenses are selected from the galaxies in the Illustris-1 simulation. We notice that, regardless of the adopted mass models (the BPL model or its special cases), the Einstein radius can be robustly determined from imaging data alone, and the median bias is typically less than $1\%$. Away from the Einstein radius, the lens mass distribution tends to be harder to measure, especially at radii where there are no lensed images detected. We find that, with rigid priors, the BPL model can clearly outperform the single power-law models by achieving $<5\%$ median bias on the radial convergence profile within the Einstein radius. As for the source light reconstructions, they are found to be sensitive to both lens light contamination and lens mass models, where the BPL model with rigid priors still performs best when there is no lens light contamination. We show that, by correcting for the projection effect, the BPL model can estimate the aperture and luminosity weighted line-of-sight velocity dispersions to an accuracy of $\sim6\%$ scatter. These results highlight the great potential of the BPL model in strong lensing related studies.
\end{abstract}

\keywords{dark matter --- galaxies: halos --- galaxies: kinematics and dynamics --- gravitational lensing: strong}

\section{Introduction}\label{sec:intro}
Strong gravitational lensing (SL) has proven to be an important tool to learn about the Universe because of its sensitivity to the geometry of the Universe and the matter distribution therein, for example, by constraining cosmological parameters \citep{2010Sci...329..924J,2012MNRAS.424.2864C,2015ApJ...806..185C,2016PhRvD..94h3510L}, testing gravity \citep{2006PhRvD..74f1501B,2009astro2010S.159K,2018Sci...360.1342C}, and measuring the mass distribution of intervening objects \citep{2008ApJ...685...70S,2012ApJ...757...22C,2020JCAP...11..045G,2022MNRAS.513.2349C}. In addition, SL can magnify the distant galaxies and help us look into their properties in more detail \citep[\eg,][]{2007ApJ...671.1196M,2011ApJ...734..104N}.

In recent years, one of the most attention-getting measurements from SL observations is the constraint on Hubble constant $H_0$ using time-delay cosmography \citep{2013ApJ...766...70S,2017MNRAS.465.4914B,2020A&A...643A.165B,2020MNRAS.494.6072S,2020MNRAS.498.1420W}. For instance, \citet{2020MNRAS.494.6072S} reported $H_0=74.2_{-3.0}^{+2.7}~\ksm$ based on the time-delay system DES J0408--5354 that has two sets of multiple images at different redshifts. The $H_0$ Lenses in COSMOGRAIL's Wellspring (H0LiCOW) collaboration found $H_0=73.3_{-1.8}^{+1.7}~\ksm$ from a joint analysis of six time-delay systems \citep{2020MNRAS.498.1420W}, in $3.1\sigma$ tension with Planck observations of the cosmic microwave background \citep{2020A&A...641A...6P}. By relaxing the assumptions on lens mass models, \citet{2020A&A...643A.165B} performed $H_0$ inference based on seven time-delay systems (one is the DES J0408--5354 and six are from H0LiCOW) along with 33 Sloan Lens ACS (SLACS) lenses selected for improving the constraints on lens mass profiles at the population level, resulting in a $\sim5\%$ measurement of $H_0$. In light of these reported precisions, we have high hopes of achieving $1\%$ precision in $H_0$ using tens of time-delay systems. 

However, there is much debate over these reported high precision of $H_0$ measurements, because there exist many lensing degeneracies making it very difficult to accurately recover the mass distribution along the line-of-sight \citep[LOS; \eg,][]{2013A&A...559A..37S, 2014A&A...564A.103S,2016MNRAS.456..739X,2018MNRAS.474.4648S,2020MNRAS.493.1725K,2020A&A...639A.101M,2021MNRAS.501.5021K}. Inaccurate reconstruction of the lensing mass distribution, especially the main lens that dominates the lensing potential along the LOS, may lead to uncontrollable systematics in $H_0$ estimation.

Among the lensing degeneracies, the most famous one is the mass-sheet degeneracy \citep[MSD;][]{1985ApJ...289L...1F}, which indicates the lensed images produced by a convergence profile $\kappa(\theta)+\kappa_{\rm ext}$ can be totally recovered by a transformed profile $(1-\kappa_{\rm ext}^\prime)\kappa(\theta)/(1-\kappa_{\rm ext})+\kappa_{\rm ext}^\prime$ with a corresponding rescaling of the source plane, where $\kappa(\theta)$, $\kappa_{\rm ext}$ and $\kappa_{\rm ext}^\prime$ represent the convergence profiles of the main lens, the actual external convergence, and a guess about $\kappa_{\rm ext}$, respectively. Since external convergence is not a direct observable, it is usually very hard to quantify with high accuracy \citep{2009ApJ...690..670T,2010ApJ...711..201S,2011ApJ...728...33G,2020MNRAS.498.1406T}, thus leading to large uncertainties in the determination of mass distribution and thus $H_0$.

The MSD is a special case of the source-position transformation \citep[SPT;][]{2014A&A...564A.103S}. The SPT refers to the fact that different combinations of lens mass and source light distributions can produce very similar or indistinguishable lensed images \citep{2013A&A...559A..37S,2017A&A...601A..77U}. \citet{2013A&A...559A..37S} illustrated the impact of SPT on the determination of $H_0$, and showed that the predicted $H_0$ can deviate by $\sim20\%$ if the lensed images produced by a composite lens are fitted by the singular power-law (SPL) model.

In addition to the MSD or SPT, there are more lensing degeneracies, such as the monopole degeneracy in regions without lensed images \citep{2000AJ....120.1654S,2012MNRAS.425.1772L} and the local degeneracies just around the lensed images \citep{1988ApJ...327..693G,2018A&A...620A..86W}. These degeneracies can bring about more uncertainties in the determination of lens mass distribution as well as $H_0$. Actually, not only $H_0$, but many other quantities relevant to SL analyses, \eg, cosmological distance ratios and parameterized post-Newtonian parameter $\gamma_{\rm PPN}$ \citep{2006PhRvD..74f1501B,2010ApJ...708..750S}, are vulnerable to the above-mentioned lensing degeneracies.

In view of these degeneracies, a question arises about what quantities can be determined faithfully by lensed images. Usually, the Einstein radius is deemed to be such a quantity \citep{SWM2006book,2010ARA&A..48...87T}. Its measurement error is typically of the order of a few percent \citep{2008ApJ...682..964B,2015ApJ...803...71S,2016ApJ...833..264S,2021MNRAS.503.2380S,2022MNRAS.517.3275E}. However, there are also works that show surprising results. Recently, based on simulated SL systems with mass distribution inferred from Sloan Digital Sky Survey (SDSS)-MaNGA stellar dynamics data \citep{2019MNRAS.490.2124L}, \citet{2022RAA....22b5014C} concluded that the Einstein radius can be recovered with $0.1\%$ accuracy. On the other hand, \citet{2018MNRAS.479.4108M} found a large scatter of $\sim20\%$ in Einstein radius estimation when comparing the singular isothermal ellipsoid (SIE) model fittings to the direct convergence fittings, where the mock lensed images are created using galaxies from EAGLE simulations \citep{2015MNRAS.446..521S}.

The other quantity, which can be constrained independently of lens mass models, is $\xi_2=R_{\rm E}{\alpha^{\prime\prime}}_{\rm E}/(1-\kappa_{\rm E})$, where ${\alpha^{\prime\prime}}_{\rm E}$ is the second-order radial derivative of the deflection profile at Einstein radius $R_{\rm E}$ and $\kappa_{\rm E}$ is the convergence at $R_{\rm E}$ \citep{2018MNRAS.474.4648S,2020MNRAS.493.1725K,2021ApJ...919...38B,2021MNRAS.501.5021K}. For axisymmetric or moderate elliptical lenses, $R_{\rm E}$ and $\xi_2$ are the two parameters that lensed images can constrain reliably. \citet{2018MNRAS.474.4648S} argued that, in order to avoid over-constraining the lens mass distribution, a lens model should have at least 3 degrees of freedom in the radial direction. The SPL model with only two radial parameters should be abandoned if higher accuracies are needed in the determination of cosmological parameters, as suggested by \citet{2020MNRAS.493.1725K}.

Nevertheless, only when lens mass distributions are reconstructed more accurately can the relevant cosmological parameters be estimated more reliably. In SL analyses, a lens model is preferred if its deflection field can be computed analytically. Many spherical models meet this requirement \citep[see a list of lens models shown in ][]{2001astro.ph..2341K}, while only a handful of elliptical models have analytic expressions for deflections, \eg, softened power-law \citep{2014MNRAS.437.1051W}, 3D broken power-law \citep[BPL;][]{2020ApJ...892...62D} and 2D BPL \citep{2021MNRAS.501.3687O} models. The commonly adopted SIE \citep{1993ApJ...417..450K,1998ApJ...495..157K} and SPL models \citep{2015A&A...580A..79T} are the special cases of the above-mentioned elliptical models.

Among the analytical lens mass models, the BPL model proposed by \citet{2020ApJ...892...62D} is a flexible model with 4 degrees of freedom in the radial direction, which can describe not only the mass distribution of lenses with a flat core, but also with a steep cusp. Furthermore, it can also fit well the well-known Navarro-Frenk-White \citep{1997ApJ...490..493N} and Einasto \citep{1965TrAlm...5...87E} profiles within sufficiently large radii. In this paper, we concentrate on this BPL model and, based on simulated SL systems, look into how accurate the lens mass distribution can be recovered by fitting the lensed images.

The rest of this paper is organized as follows. In Section \ref{sec:bpl}, we briefly review the basics about BPL model. In Section \ref{sec:mock}, we show the creation of mock lensing observations, which are used to evaluate the BPL model as well as the SIE and SPL models. We describe in detail the extraction and reconstruction of lensed images in Section \ref{sec:recon} and investigate the necessary priors in lens mass modeling. Results are presented in Section \ref{sec:result}. Conclusion and discussions are given in the last section.

\section{The BPL model}\label{sec:bpl}
In this section, we briefly review the basics about the BPL model, including its density profile and deflections. We also show the formalism for estimating the line-of-sight velocity dispersions (LOSVDs) observed by single-fiber spectroscopy, \ie, the aperture and luminosity (AL) weighted LOSVDs, which provide importantly the dynamical information for calibrating the lensing mass distribution.

\subsection{The volume and surface density profile}
The volume density profile of BPL model is expressed by
\begin{equation}\label{eq:rho}
\rho(r)=
\begin{cases}
\rho_c\left(r/r_c\right)^{-\alpha_c} & \text{if $r\leqslant r_c$}\\
\rho_c\left(r/r_c\right)^{-\alpha} & \text{if $r\geqslant r_c$},
\end{cases}
\end{equation}
where $\alpha_c$ and $\alpha$ are, respectively, the inner and outer slopes, and $\rho_c$ is the volume density at break radius $r_c$.

By integrating $\rho(r)$ along the LOS, one obtains the surface mass density profile $\Sigma(R)$, and then has the convergence profile $\kappa(R)=\Sigma(R)/\Sigma_{\rm crit}$ with
\begin{equation}\label{eq:sigc}
\Sigma_{\rm crit}=\frac{c^2}{4\pi G}\frac{D_s}{D_dD_{ds}}
\end{equation}
known as the critical surface mass density in lensing analyses, where $D_d$, $D_s$, and $D_{ds}$ are the angular diameter distances to the lens deflector, to the source, and from the deflector to the source, respectively.

As shown by \citet{2020ApJ...892...62D}, $\kappa(R)$ can be expressed in terms of two parts as
\begin{equation}\label{eq:kappa12}
\kappa(R)=\kappa_1(R)+\kappa_2(R),
\end{equation}
with
\begin{equation}\label{eq:kappa1}
\kappa_1(R)=\frac{3-\alpha}{2}\left(\frac{b}{R}\right)^{\alpha-1}
\end{equation}
corresponding to a single power-law part and
\begin{align}\label{eq:kappa2}
\kappa_2(R)=&\frac{3-\alpha}{ \mathcal{B}(\alpha)}\left(\frac{b}{r_c}\right)^{\alpha-1} \nonumber\\
{}&\times\tilde{z}\left[F\left(\frac{\alpha_c}{2},1;\frac{3}{2};\tilde{z}^2\right)-F\left(\frac{\alpha}{2},1;\frac{3}{2};\tilde{z}^2\right)\right]
\end{align}
a complementary part within $r_c$, which is a mass deficit for $\alpha_c<\alpha$ or a mass surplus for $\alpha_c>\alpha$, where $\mathcal{B}(\alpha)={\rm Beta}\left(\frac{1}{2},\frac{\alpha-1}{2}\right)$, $\tilde{z}=\sqrt{1-R^2/r_c^2}$, $F()$ denotes the Gauss hypergeometric function, and $b$ is a scale radius defined by
\begin{equation}\label{eq:brhoc}
b^{\alpha-1}=\frac{\mathcal{B}(\alpha)}{\Sigma_{\rm crit}}\frac{ 2}{3-\alpha}\rho_c r_c^\alpha .
\end{equation}
Note that $\kappa_2(R)$ is zero when $R \geqslant r_c$ or $\alpha_c=\alpha$.

In order to describe surface mass distribution that is elliptically symmetric, the circular radius $R$ in $\kappa(R)$ can be generalized to elliptical radius $R_{\rm el}=\sqrt{qx^2+y^2/q}$, where $q$ is the axis ratio of the elliptical isodensities. In this case, the area enclosed by $R_{\rm el}$ is $\pi R_{\rm el}^2$ independent of $q$. In view of this advantage, we thus can define an effective Einstein radius of a lens as the elliptical radius within which the mean convergence is unity.

Central black holes can have detectable effects on the formation of central images \citep{2001MNRAS.323..301M} and be closely related to the velocity dispersions \citep{2000ApJ...539L..13G}. In this paper, we adopt the simulated galaxies from the Illustris-1 simulation that accounts for the formation of black holes, as mock lenses (see Section \ref{sec:mock}). So, when modeling the mass distribution of a lens, we take the effect of a central black hole into consideration. Also note that the elliptical BPL model becomes the SPL model when $\alpha_c=\alpha$ and the SIE model when $\alpha_c=\alpha=2$.

\subsection{The deflection angles}
In complex notation, the lens equation is
\begin{equation}\label{eq:lenseq}
  z_s=z-\alpha(z),
\end{equation}
which relates the true source position $z_s=x_s+{\rm i}y_s$ on source plane to its observed position $z=x+{\rm i}y$ on lens plane (please do not confuse the complex numbers $z$ in this subsection with the redshift symbols in other sections), where $\alpha(z)=\alpha_x+{\rm i}\alpha_y$ is the scaled deflection angle caused by the intervening lens \citep{SWM2006book,2015A&A...580A..79T}. If the lens mass distribution is elliptically symmetric, the deflection angle $\alpha(z)$ at position $z$ can be evaluated by
\begin{eqnarray}\label{eq:alp_star}
\alpha^*(z)&=&\frac{2}{z}\int_0^{R_{\rm el}}\frac{\kappa(R)R \ \dd R}{\sqrt{1-\zeta^2 R^2}},
\end{eqnarray}
where the symbol $*$ denotes the complex conjugate, $\zeta^2=(1/q-q)/z^2$ and $R_{\rm el}$ is the elliptical radius of the ellipse passing through the position $z$ \citep{1975ApJ...195...13B,1984MNRAS.208..511B}. In the special case of $q=1$, we can easily find $\alpha^*(z)=R^2\bar{\kappa}(R)/z$. Thus, for a point mass, \eg, a black hole with mass $m_b$, its deflection can be expressed by
\begin{eqnarray}\label{eq:alp_bh}
{\alpha_b}^*(z)&=&\frac{1}{\pi}\frac{m_b}{\Sigma_{\rm cirt}}\frac{1}{z}.
\end{eqnarray}

For the BPL model, by substituting Equation (\ref{eq:kappa12}) into Equation (\ref{eq:alp_star}), we have the deflections $\alpha^*(z)={\alpha_1}^*(z)+{\alpha_2}^*(z)$ with
\begin{equation}
{\alpha_1}^*(z)=\frac{R_{\rm el}^2}{z}\left(\frac{b}{R_{\rm el}}\right)^{\alpha-1}F\left(\frac{1}{2},\frac{3-\alpha}{2};\frac{5-\alpha}{2};\zeta^2 R_{\rm el}^2\right)
\end{equation}
for the power-law part $\kappa_1$ and
\begin{align}
{\alpha_2}^*(z)=&\frac{r_c^2}{z}\frac{3-\alpha}{\mathcal{B}(\alpha)}\left(\frac{b}{r_c}\right)^{\alpha-1}
\left[\frac{2}{3-\alpha_c}\mathscr{F}\left(\frac{3-\alpha_c}{2},\mathcal{C}\right)-\right. \nonumber\\
{}&\left.\frac{2}{3-\alpha}\mathscr{F}\left(\frac{3-\alpha}{2},\mathcal{C}\right)-
  \mathcal{S}_0(\alpha,\alpha_c,\tilde{z}_{\rm el},\mathcal{C})\right]
\end{align}
for the complementary part $\kappa_2$, where $\mathcal{C}=r_c^2\zeta^2$, $\tilde{z}_{\rm el}=\sqrt{1-R_{\rm el}^2/r_c^2}$, and functions $\mathscr{F}$ and $\mathcal{S}_0$ can be written in terms of the Gauss hypergeometric functions. \footnote{In this paper, the Gauss hypergeometric function is computed using the function scipy.special.hyp2f1 in Python, where the latest version of SciPy library is recommended since it solves the divergence problem of function hyp2f1 in some regions.} Please refer to \citet{2020ApJ...892...62D} for detailed expressions of $\mathscr{F}$ and $\mathcal{S}_0$. Note that $\mathcal{S}_0$ disappears when $R_{\rm el}\geqslant r_c$ or $\alpha_c=\alpha$.

\subsection{The AL-weighted LOSVDs}\label{subsec:losvd}
In addition to testing the accuracy of mass measurements of galaxy-scale lenses, we are also interested in the accuracy of AL-weighted LOSVDs predicted from the reconstructed mass distribution, which can help us calibrate the lens mass distribution.

Based on the spherical Jeans equation, the AL-weighted LOSVD for a galaxy with constant velocity anisotropy can be modeled as
\begin{equation}\label{eq:al_losvd1}
\langle\sigma_\parallel^2\rangle=\frac{\int_0^\infty \dd R~ R w(R)\int_{-\infty}^\infty \dd Z~ j(r)(1-\beta R^2/r^2)\sigma_r^2(r) }{\int_0^\infty \dd R~ R w(R)\int_{-\infty}^\infty \dd Z~ j(r)}
\end{equation}
where $\sigma_r^2(r)$ is the radial velocity dispersion for stars, $\beta$ denotes the global velocity anisotropy, $j(r)$ is the 3D luminosity density profile, and $w(R)$ is a weighting function accounting for the fiber size and seeing effect for ground-based spectroscopic observations.

By assuming $w(R)$ follows a Gaussian distribution \citep{2010ApJ...708..750S}, \ie,
\begin{equation}\label{eq:gauwr}
w(R)\approx\exp\left(-\frac{R^2}{2\sigma_{\rm fib}^2}\right),
\end{equation}
where $\sigma_{\rm fib}$ is a function of seeing and fiber size, \citet{2020ApJ...892...62D} found that Equation (\ref{eq:al_losvd1}) can be transformed into
\begin{eqnarray}\label{eq:allosvd}
&&\langle\sigma_\parallel^2\rangle= \nonumber \\
&&\frac{\int_0^\infty \dd r~ r^2 j(r)\sigma_r^2(r)
\left[\Phi\left(1,\frac{3}{2};-\frac{r^2}{2\sigma_{\rm fib}^2}\right)-\frac{2\beta}{3}\Phi\left(2,\frac{5}{2};-\frac{r^2}{2\sigma_{\rm fib}^2}\right) \right] }{\int_0^\infty \dd r~ r^2 j(r) \Phi\left(1,\frac{3}{2};-\frac{r^2}{2\sigma_{\rm fib}^2}\right)} \nonumber \\
\end{eqnarray}
with only 1D integrals, where $\Phi(a_1,a_2;-x)$ is the Kummer's confluent hypergeometric function.

We adopt the power-law S\'ersic (PL-S\'ersic) profile to describe the light (or stellar mass) density profile \citep{2005MNRAS.362..197T}, which is written as
\begin{equation}\label{eq:jplsersic}
j(r)=\left\{
           \begin{array}{ll}
             \displaystyle j_c\left(r/r_c\right)^{-\alpha_c} &
\hbox{if $r\leqslant r_c$} \\\\
             \displaystyle j_0\left(\frac{r}{s}\right)^{-u}\exp\left[-\left(\frac{r}{s}\right)^\nu\right]
& \hbox{if $r\geqslant r_c$}
           \end{array}
         \right.
\end{equation}
where
\begin{equation}
j_0=j_c\left(\frac{r_c}{s}\right)^{u}\exp\left[\left(\frac{r_c}{s}\right)^\nu\right] ,
\end{equation}
$j_c$ is the luminosity density at break radius $r_c$, $s=R_{\rm eff}/k^n$ is a scale radius defined by the 2D effective radius $R_{\rm eff}$ for the S\'ersic profile and the S\'ersic index $n$ ($k$ here is a function of $n$ and its expression can be found in \citealt{1999A&A...352..447C} and \citealt{2003ApJ...582..689M}), $\nu =1/n$, and $u=1-0.6097\nu+0.054635\nu^2$ \citep{1999MNRAS.309..481L,2001A&A...379..767M}.

Given the BPL mass model and PL-S\'ersic light profile, the radial velocity dispersion $\sigma_r^2(r)$ can be analytically calculated. Please refer to the Equations (41)--(45) in \citet{2020ApJ...892...62D} for the analytical expressions of $\sigma_r^2(r)$, where the PL-S\'ersic profile is assumed to have the same break radius $r_c$ and inner slope $\alpha_c$ as the BPL model.

\section{Mock lensing observations}\label{sec:mock}
In this section, we first display some basics about observed SL systems, which are used for reference, and then briefly describe the Illustris simulation from which a sample of galaxies are selected as mock lenses. In the third subsection, we show the generation of hundreds of mock SL systems with single or multiple exposures.

\subsection{Observational data}
In order to produce realistic lensing images, we refer to the lenses detected by the Sloan Lens ACS (SLACS) and the SLACS for the mass (S4TM) surveys \citep{2008ApJ...682..964B,2015ApJ...803...71S}. The SL candidates for these two surveys are first identified by taking advantage of the galaxy spectra from SDSS. Then, follow-up imaging observations are performed with the Advanced Camera for Surveys (ACS) aboard the Hubble Space Telescope (HST).

By visually inspecting the high-resolution HST images, the SL candidates are classified into different classes, where the ones with clear and definite lensing features are termed ``grade-A'' lenses. In the following analyses, we use the grade-A systems with only one dominant lens for reference. In total, we have 63 SLACS lenses (20 with one exposure and 43 with multiple exposures) and 38 S4TM lenses (all with one exposure), most of which are elliptical galaxies. Hereafter, we refer to both SLACS and S4TM surveys simply as SLACS surveys.

\begin{figure*}
  \centering
  \includegraphics[width=0.9\textwidth]{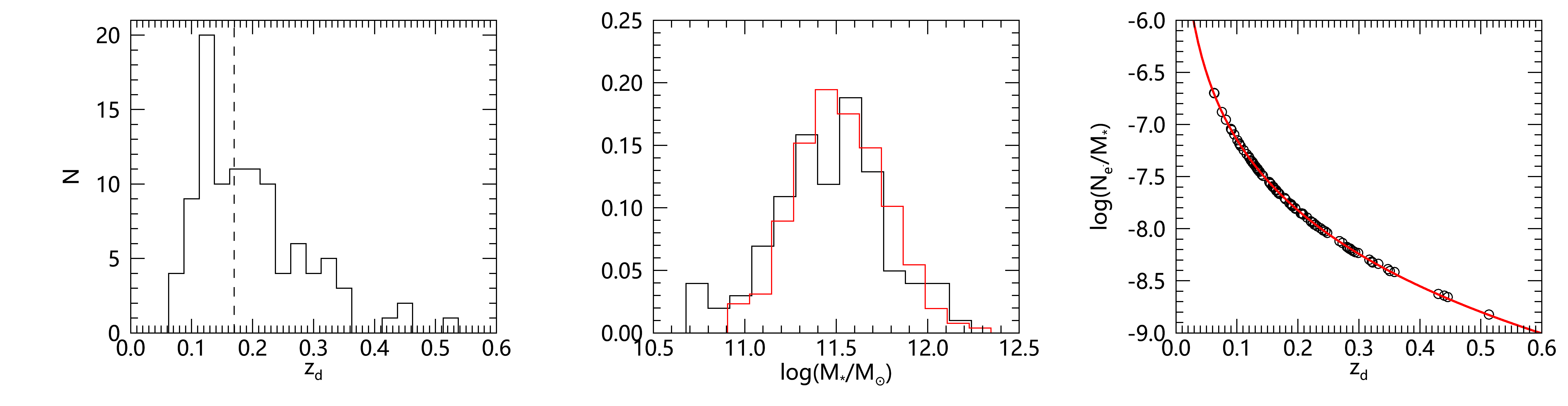}\\
  \caption{Some basic information about 101 reference lenses (63 SLACS and 38 S4TM lenses). Left: the distribution of lens redshift $z_d$. The vertical dashed line indicates the median redshift of the lenses. Middle: the normalized stellar mass function of the 101 real lenses (black histogram) and the 257 ``grade-A'' mock lenses (red histogram). Right: the ratio of $N_e$ and $M_\star$ as a function of redshift, where the red line corresponds to the prediction from Equation (\ref{eq:nemstar}). Each circle corresponds to a lens with total light intensity $N_e$ and total stellar mass $M_\star$, where $N_e$ is estimated from the core-S\'ersic profile fit to the HST image (in electrons per second), and $M_\star$ is estimated by matching the single stellar-population models with Chabrier initial mass function to its AB magnitude \citep{2015ApJ...803...71S}.}\label{Fig:obszmass}
\end{figure*}

Figure \ref{Fig:obszmass} illustrates some basic information about the 101 SLACS lenses. The left panel shows the redshift distribution of the lenses. The middle panel presents the stellar mass distribution of the SLACS (black histogram) and mock lenses (red histogram). The stellar masses of the SLACS lenses are estimated by scaling single stellar-population models (assuming the Chabrier stellar initial mass function that is also used in the Illustris simulations) to fit their HST F814W photometry \citep{2015ApJ...803...71S}. We thus expect that a scaling relation may exist between the apparent magnitude and the stellar masses of these lenses. The last panel in Figure \ref{Fig:obszmass} shows the ratio of the total light intensity $N_e$ (determined by the observational data in units of electrons per second) to the estimated stellar mass $M_\star$ as a function of lens redshift. We find that the $N_e$ is related to $M_\star$ for the SLACS lenses as
\begin{equation}\label{eq:nemstar}
N_e\approx\frac{0.2M_\star}{4\pi D_L^2(1+z)^{0.5}} ,
\end{equation}
where $D_L$ is the luminosity distance in units of megaparsecs, and $M_\star$ is in units of $\msun$. Using this relation, we can easily convert stellar masses of simulated galaxies to light intensities to mimic HST-like observations.

\subsection{The Illustris simulation}
The Illustris project runs a set of large-scale hydrodynamic simulations by accounting for various kinds of baryonic physics including gas cooling, star formation and evolution, and feedback from active galactic nuclei and supernovae \citep{2014MNRAS.445..175G,2014MNRAS.444.1518V,2015A&C....13...12N}. It assumes a flat $\Lambda$ cold dark matter cosmology with $\Omega_m=0.2726$, $\Omega_\Lambda=0.7274$, $\Omega_b=0.0456$, $\sigma_8=0.809$, $n_s=0.963$, and $h=0.704$. Among the simulation runs, the one with the highest resolution is named as the Illustris-1 simulation from which we extract massive galaxies to generate mock SL systems.

Specifically, the Illustris-1 simulation follows $1820^3$ dark matter particles and initially $1820^3$ hydrodynamical cells in a periodic box of size $75~\hmpc$. The mass for a dark matter particle is $\sim6.26\times10^6~\msun$, and the baryonic matter has an initial mass resolution of $\sim1.26\times10^6~\msun$. At the end of the simulation, the minimum mass for a baryonic matter particle reaches to a few times $10^4~\msun$. The simulation assigns different softening lengths to different types of particles. For example, the softening length is fixed to be a comoving value of 1 \hkpc for dark matter particles and is larger than 0.5 \hkpc in physical scale for stars and black holes. For gas cells, an adaptive softening length is defined according to the fiducial cell size, with a minimum equal to that used for the collisionless baryonic particles. 

\subsection{Generation of mock observations} \label{subs:mockobs}
Using the Illustris-1 simulation at redshift zero, we extract 5343 galaxies with stellar mass larger than $10^{10}~h^{-1}\msun$. These galaxies are artificially put at redshift $z_d\simeq0.178$ which is close to the median redshift of the SLACS lenses (see the vertical dashed line shown in the left panel of Figure \ref{Fig:obszmass}). At this redshift, $1^{\prime\prime}$ corresponds to $3~\kpc$ for the cosmology adopted by the Illustris project.

Based on thin lens approximation, we project the 3D mass distribution of the galaxies along the $x$-direction of the simulation box onto the 2D lens plane at $z_d\simeq0.178$ to analyze their lensing effect. In the process of projection, all of the matter components are considered including the dark matter, stars, gas, and black holes. We assume the source plane is at redshift $z_s=0.6$, which is around the median redshift of the background sources of the SLACS lensing systems. In this case, the critical surface mass density is $\Sigma_{\rm crit}\simeq4.0\times10^{15}\msun/\mpc^{2}$ for the Illustris-adopted cosmology.

We define the ``true'' Einstein radius for each galaxy as the radius of a circle within which the mean convergence is unity. Based on these true Einstein radii, we find that only a small fraction of galaxies can produce noticeable lensed images. For example, there are only $\sim350$ galaxies whose Einstein radii are larger than $0\farcs5$. With the purpose of generating SL systems whose statistics are similar to those of the SLACS survey, we choose 283 galaxies with Einstein radii in the range of $0\farcs6$--$2^{\prime\prime}$ as the main sample for further analyses, of which about 85\% are elliptical galaxies.

Based on the 283 galaxies, we first pixelize their projected overall and stellar mass distribution using the triangular shaped cloud algorithm \citep{1981CSUP}, respectively, with a resolution of $0\farcs05$. We then get 283 convergence maps by scaling the projected overall mass density distributions with $\Sigma_{\rm crit}$. By resorting to Equation (\ref{eq:nemstar}) and assuming a constant mass-to-light ratio, the light distribution in units of electrons per second can thus be derived for the lenses from their stellar mass distribution.

The deflection angle can be expressed as a convolution product between the convergence map $\kappa(\bm{x})$ and the kernel $\bm{x}/|\bm{x}|^2$. Given a surface mass distribution sampled in a regular grid, it is easy to derive the deflection field in Fourier space by applying the convolution theorem. For the 283 galaxy lenses, their deflection angles at grid points are thus evaluated in this way, where the field of view (FOV) of the convergence map for each lens is large enough to cover its entire halo with zero-padding. By tracing the light rays back to the source plane based on the lens equation, we can construct lensed images of background sources. SL images without photometric noise can thus be obtained by directly adding together the light distribution of foreground galaxies and lensed images.

In this paper, we assume each background source is a faint and compact galaxy which follows a 2D S\'ersic profile $I(R)=I_0\exp\left[-k\left(\frac{R}{R_{\rm eff}}\right)^\nu\right]$ with $I_0=0.5~\rm e^- s^{-1} pixel^{-1}$, $R_{\rm eff}=0\farcs15$ (about 1~\kpc at redshift $z_s=0.6$), and S\'ersic index $n=1$. In this case, the corresponding apparent magnitude is $\sim 23.5$ for the background sources. The properties of source galaxies considered here are roughly consistent with those found by \citet{2011ApJ...734..104N}.

More specifically, the position angle of each source galaxy is randomly assigned in the range of 0--$\pi$ and its axis ratio is in the range of 0.2--1. We randomly place a source galaxy on the source plane around the center of the foreground lens with a scatter of $0\farcs2$ in both $x$ and $y$-directions. A lensed image will be a grade-A candidate if its magnification is larger than 5 and the number of pixels in the feature mask is more than 500, where the feature mask is defined as the pixels whose light intensities are 5\% brighter than those of the lens itself. We notice that the feature mask defined in this way is roughly consistent with (or somewhat larger than) the region visually identified for the lensing features (see the gray shaded area shown in the left panels of Figure \ref{Fig:bspex} for examples of feature masks). In practice, we cannot determine the feature mask so straightforwardly because we do not know exactly the true lensed images due to the existence of lens light contamination.

With the aim of producing more realistic lensing images, photometric noises are considered, \eg, a sky background of $0.11~\rm e^- s^{-1}$ and the Poisson noise on each pixel. The effect of the point spread function (PSF) is also taken into account by randomly applying the PSF in SLACS images to mock images. Bad pixels due to cosmic rays are added according to the bad pixel distribution in HST images. We also account for a readout noise of $\rm 5~e^-$ for each pixel. By default, the exposure time is 420~s for each lensing image, corresponding to the single exposure time of the SLACS survey. For comparison, we also generate SL images with four exposures (2200~s in total). The left panels of Figure \ref{Fig:bspex} show two examples of mock SL images with a single exposure.

It should be mentioned that, owing to the existence of large flat cores of Illustris galaxies arising from the softening effect in numerical simulations, most of the pure lensed images in our mock catalog exhibit lensing features in or toward the lens centers. This may not be the case for real SL images. Furthermore, due to the contamination of foreground lens light, the central images may disappear or hardly be detected in the lensing images. The missing of central images may preclude the accurate measurement of inner density profile.

Based on the kinematics of stellar particles, \citet{2020ApJ...892...62D} have calculated the AL-weighted LOSVDs for the Illustris galaxies adopted here, where the galaxies are assumed to be at redshift $z_d=0.178$ and observed by SDSS-like spectroscopy with a fiber radius of $1\farcs5$ and in a seeing condition of $1\farcs69$. We take the AL-weighted LOSVDs evaluated by \citet{2020ApJ...892...62D} directly as true values in the following analyses.

\section{Lensing mass reconstruction}\label{sec:recon}
This section includes two parts. One is for the subtraction of foreground lens light distribution using the B-spline technique, and the other is for the reconstruction of lensed images, where priors on mass distribution are investigated in detail.

\subsection{Subtraction of lens galaxy light}
For most of galaxy-scale lenses, their Einstein radii are typically smaller than their half-light radii, making it difficult to clearly identify the relatively faint lensed images. In order to subtract the foreground light distribution, parametric light profiles, \eg, S\'ersic or double S\'ersic profiles \citep{2008ApJ...682..964B,2019MNRAS.483.5649S,2020A&A...643A.165B,2022MNRAS.517.3275E}, are usually adopted as fitting models. However, these profiles with a limited number of parameters may result in undesired residuals, especially for the light distribution with complex angular structures.

In this paper, we use the B-spline technique to fit the foreground light distribution \citep{2001pgts.book.....D,2006ApJ...638..703B}. The B-spline fitting model can be written as
\begin{equation}
I_{\rm bsp}(R_{\rm el},\theta)=\sum_{m,k}\left[b_{mk}\cos (m\theta)+c_{mk}\sin(m\theta)\right]f_k(R_{\rm el}),
\end{equation}
where $f_k(R_{\rm el})$ are the radial basis functions with elliptical radius $R_{\rm el}$, and $m$ denotes the multipole orders in the $\theta$-direction. $b_{mk}$ and $c_{mk}$ are the coefficients that can be determined by fitting the light distribution in a sense of least-squares.

For the B-spline technique, the lensing features need to be estimated and masked in advance so as to get unbiased fittings to the lens light distribution (see Section \ref{subs:mockobs} for the definition of feature mask). By masking the estimated lensing features, we then fit the monopole term (\ie, $m=0$) of light distribution to estimate their centers, ellipticities, and position angles. With the centers, ellipticities, and position angles fixed, the light distribution is fitted again by adding higher order even multipoles with $m=[2,4,6,8]$. If there are obvious odd features left in the residual for a lens after subtracting the even multipoles, we will further add the odd terms with $m=[1,3,5,7]$ to fit the lens light distribution once more. If the extracted lensing features cannot be clearly recognized no matter what types of multipole terms are included, the corresponding lens systems will be discarded.

Finally, we have 257 ``grade-A'' lenses left for the single exposure catalog, where 18 lenses show more clear lensing features if the odd multipole terms are included in the B-spline fittings. For the mock catalog with four exposures, we almost have the same cases left, indicating that more exposures may not help us identify more ``grade-A'' galaxy-scale lenses, because of the significant contamination from foreground lens light. We thus use these 257 lenses with both single and multiple exposures in the following analyses.

\begin{figure*}
  \centering
  \includegraphics[width=0.95\textwidth]{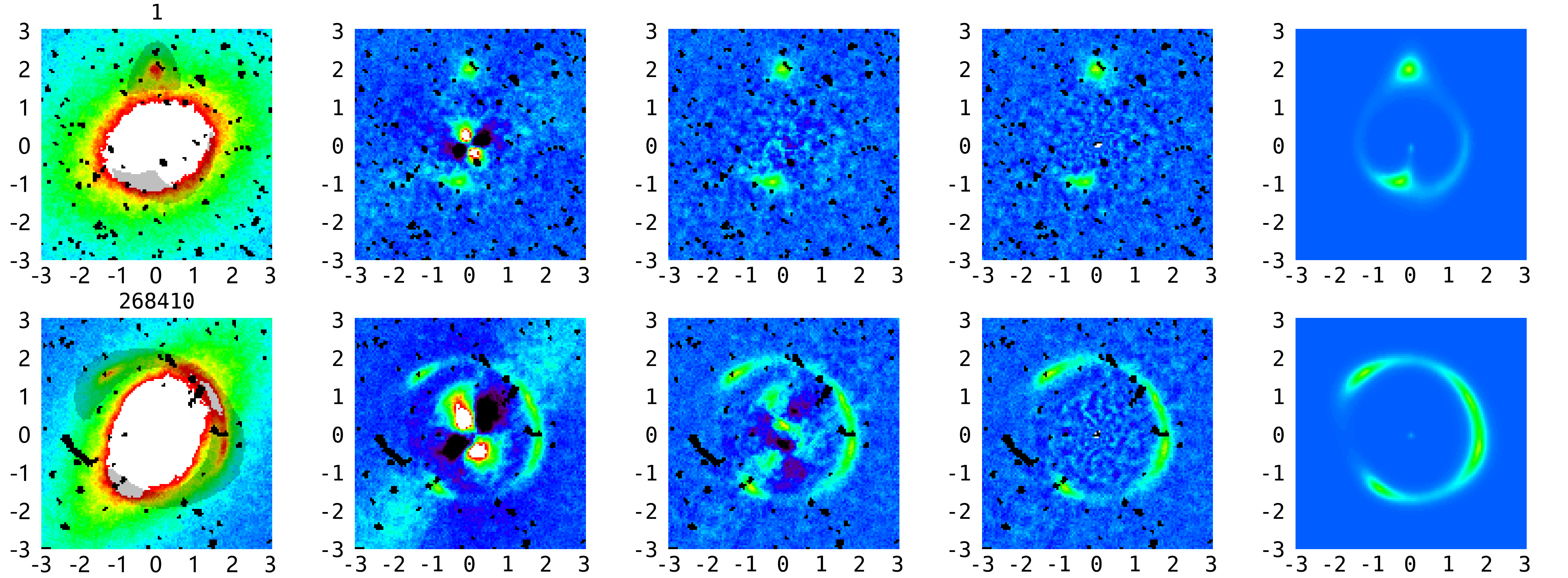}\\
  \caption{Two examples of extracting lensing features based on the B-spline technique. Shown here for each lens is the central region in the FOV of $6''\times6''$. The top and bottom rows are for the lenses with $\rm{id}=1$ and 268410, respectively. The left panels show the lensing images with lens light distribution. The shadows mark the estimated region of lensed images. The second and third columns present the residual images after subtracting the B-spline fittings with monopole and even multipole terms, respectively. The fourth column shows the residual images by further subtracting the possible odd multipole terms. The rightmost column displays the intrinsic lensing features convolved with a corresponding point-spread function (PSF). Note that, the resolution for all of these maps is $0\farcs05$, and the bad pixels are caused by cosmic rays.}\label{Fig:bspex} 
\end{figure*}

Figure \ref{Fig:bspex} shows two examples of B-spline fittings to the single exposure images. The first column displays the SL images with foreground lens light distribution. We note that, for galaxy-scale lenses, the foreground lens light may blur the lensed images significantly and make it difficult to predefine the feature mask. The second column shows the residuals after subtracting the monopole ($m=0$) term of light distribution. We can see obvious angular structures for these two lenses. By further subtracting the even terms, as shown in the third column, angular structures almost disappear for the first lens. However, there still exist noticeable odd features for the second lens, which can be largely reduced by further accounting for the odd terms in the B-spline fitting, as demonstrated by the fourth column. For reference, we show in the last column the true lensing features convolved with PSF. We found that, for these two examples, the extracted lensing features are fairly consistent with the true ones.

\subsection{Reconstruction of lensed images}
\subsubsection{$\chi^2$ definition}
In this paper, we use the forward modeling to reconstruct the lensed images, \ie, modeling the lens mass distribution by the BPL model and the source light distribution by the S\'ersic profile. For comparison, the SIE and SPL lens mass models are also investigated.

The $\chi^2$ for the fittings to the extracted lensing features is defined according to
\begin{equation}\label{eq:chi2scale}
\chi^2=\frac{\sum_{i=1}^N(I_{\rm data}-I_{\rm bsp}-I_{\rm model})_i^2w_i^2}{\frac{1}{N^\prime}\sum_{j=1}^{N^\prime}(I_{\rm data}-I_{\rm bsp})_j^2w_j^2} ,
\end{equation}
where $I_{\rm data}$, $I_{\rm bsp}$ and $I_{\rm model}$ denote the light distribution of ``observed'' lensing image, the B-spline fitting to the foreground light, and the modeling of the lensed image convolved with PSF, respectively. The index $i$ indicates the $i$th pixel in a suitable FOV, and $j$ is for the pixels adopted for the B-spline fittings that do not take into account pixels in feature mask. The $w$ is the reciprocal of the Poisson noise error, which is proportional to the square root of the total exposure time in a pixel and is very close to the Gaussian error due to the sufficiently long exposures. Note that, for single exposure images, null weights are assigned to the bad pixels. In order to avoid possible over- or under-fittings of the B-spline technique, the denominator in Equation (\ref{eq:chi2scale}) is applied to scale the weight for each pixel systematically and to make the $\chi^2$ definition reasonable.

We employ the Python module emcee, an affine invariant Markov Chain Monte Carlo (MCMC) Ensemble sampler, to estimate the optimal values of parameters \citep{2013PASP..125..306F}. The natural logarithm of the posterior probability used for emcee is defined to be
\begin{equation}\label{}
\ln \mathcal{P}(\Theta|I_{\rm data})=-\frac{1}{2}\chi^2+\ln p(\Theta)+{\rm Const.}
\end{equation}
where $p(\Theta)$ is the prior function for a set of parameters.

In order to avoid possible local maxima in parameter space, we run the MCMC samplings twice. The first run aims to find an initial guess of the parameters for the second run, which are taken to be the values resulting in the maximum posterior probability in the first run. For both runs, a large enough number of burn-in steps with 500 walkers are set up to make sure the remaining samplings are in equilibrium. After the second run, we finally have 200,000 acceptable values saved for each parameter.

\subsubsection{Priors}
As reviewed in Section \ref{sec:intro}, there are many lensing degeneracies, which can lead to large uncertainties in the measurement of lens mass distribution but leave the lensed images and flux ratios almost invariant. The lensed images can provide only limited information about the lens mass distribution in the annulus encompassing them. In regions without clear lensing features, the estimation of the mass distribution is just an interpolation or extrapolation based on a lens mass model \citep{2020MNRAS.493.1725K}. When we choose a lens mass model, for example, with only one or two free parameters in the radial direction, rigid priors have already been applied, and the significance of systematics strongly depends on the consistency between the adopted lens model and the true mass distribution.

The BPL profile has more than 3 degrees of freedom in the radial direction. In principle, it is likely to result in more reasonable constraints on lens mass distribution. However, we find in the next section that, if we do not impose priors on the radial profile of the BPL model, the biases or systematic uncertainties for the mass distribution within Einstein radius could still be significant, depending on the configuration and signal-to-noise ratio of the extracted lensed images. Therefore, it is essential to add reasonable priors to certain parameters. In this work, we pay special attention to the priors on the central black hole mass $m_b$, axis ratio $q$, and the inner density profile ($r_c$ and $\alpha_c$) for the BPL model. These quantities are constrained by using stellar mass (or light) distribution.

\begin{figure*}
  \centering
  \includegraphics[width=0.9\textwidth]{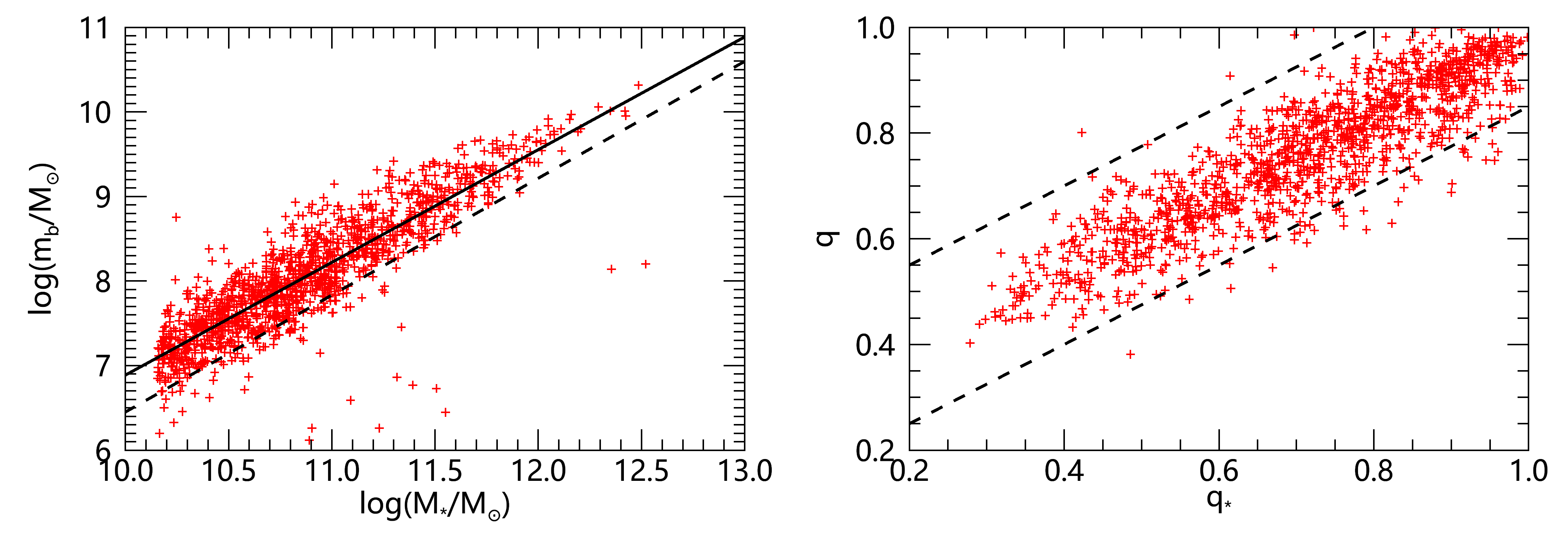}\\
  \caption{Statistical relations about the properties of 5343 Illustris-1 galaxies with stellar mass larger than $10^{10}~h^{-1}\msun$. For clarity, only the data points for $\sim 1300$ elliptical galaxies are presented. Left: relation between central black hole mass $m_b$ and total stellar mass $M_\star$. The black solid line shows the best-fit to the red pluses corresponding to the elliptical galaxies. For reference, the dashed line shows the best fit for disk galaxies. Right: comparison between axis ratios of overall mass and stellar mass distribution. The dashed lines, which confine about $95\%$ of galaxies, indicate the defined upper and lower limits of the axis ratio $q$ for overall mass distribution given the axis ratio $q_\star$ for stellar mass distribution.}\label{Fig:priors}
\end{figure*}

Due to the lack of central images or the low signal-to-noise ratio of lensed images in the central region, it is usually very hard to determine the mass of central black holes based solely on SL observations. To account for the effect of central black holes, we resort to the relation between central black hole mass $m_b$ and total stellar mass $M_\star$ of galaxies
\citep{2004ApJ...604L..89H,2013ARA&A..51..511K,2015ApJ...813...82R}, which is parameterized by
\begin{equation}\label{eq:mbms}
\log(m_b/\msun)=a+b\log(M_\star/10^{11}\msun) .
\end{equation}
The left panel of Figure \ref{Fig:priors} illustrates the fittings to the relation between $m_b$ and $M_\star$ of 5343 Illustris galaxies. The solid and dashed lines are for the elliptical and disk galaxies, respectively. For clarity, we only plot the data points for the elliptical galaxies.

Note that the galaxy types are defined according to the S\'ersic indices and stellar dynamical properties of galaxies \citep[please refer to the details in][]{2020ApJ...892...62D}. We find that, for elliptical galaxies, the best-fit values are $a=8.22$ and $b=1.33$, comparable with the observations \citep[\eg,][]{2015MNRAS.452..575S}. For disk galaxies, $a=7.83$ and $b=1.38$. The intrinsic scatters around the best-fit lines are about $0.34$ and $0.29$ dex for elliptical and disk galaxies, respectively. It should be mentioned that, in practical fitting, we do not take into account the scatter of the black hole mass for each galaxy but fix it to be the value at the best-fit relation.

The second nonnegligible quantity is the ellipticity, which by inspection has evident degeneracy with other lens model parameters, \eg, amplitude $b$ and density profile slopes \citep[see also Figure B.1 in][for example]{2020A&A...639A.101M}. In the right panel of Figure \ref{Fig:priors}, we present the correlation of axis ratios between overall mass and light (\ie, stellar mass) distributions for elliptical galaxies. For disk galaxies, we find a very similar scatter diagram with only a slight shift. These axis ratios are estimated by simultaneously fitting the corresponding surface mass and light distributions in an FoV of $14''\times14''$ using the BPL and PL-S\'ersic profiles with the same inner density slope and break radius. The two dashed lines show the upper ($q=0.75q_\star+0.4$) and lower ($q=0.75q_\star+0.1$) limits of the axis ratio $q$ of overall mass distribution given the axis ratio $q_\star$ of stellar mass distribution, which are adopted to confine the ellipticity of lens mass distribution in model fittings.

The third prior is about the inner density profile, which is hard to constrain if there are no identifiable lensed images in the inner or central region. In order to make the lens mass modeling more efficient and accurate, we propose constraining the inner part of mass distribution by resorting to light distribution, in view of the fact that baryonic matter dominates the overall mass in the core region of galaxies. For implementing this idea, we prefer using the PL-S\'ersic profile to fit the lens light distribution first, where the break radius $r_c$ is not a free parameter but is a radius at which the logarithmic slope for the S\'ersic part is specified. Then, we assume that the lens mass and light distributions share the same break radius and inner slope.

Specifically, based on Equation \ref{eq:jplsersic}, we can obtain the right derivative of the PL-S\'ersic profile at $r_c$ in logarithmic space according to
\begin{equation}
\alpha_j=-\frac{d\ln j(r)}{d\ln r}\Big|_{r=r_c}=u+\nu\left(\frac{r_c}{s}\right)^\nu .
\end{equation}
So, the break radius can be expressed by
\begin{equation}\label{eq:breakradius}
r_c=s\left[(\alpha_j-u)n\right]^n .
\end{equation}
This indicates that, for a lens light distribution, a break radius can be determined by the S\'ersic part of the PL-S\'ersic profile fitting with a given value of $\alpha_j$. For instance, even for a pure S\'ersic profile, we may also find a break radius at which the logarithmic slope is $\alpha_j$ and within which the density profile is assumed to follow a power law.

So, the aim now is to find a break radius for each lens using the PL-S\'ersic profile with a given value of $\alpha_j$ to fit the lens light distribution. By analyzing the light distribution of Illustris-1 galaxies, we find that the break radius $r_{c,2.3}$ defined by $\alpha_j=2.3$ is a fairly good choice for distinguishing the inner part from the outer part of mock lenses.

\begin{figure*}
  \centering
  \includegraphics[width=0.9\textwidth]{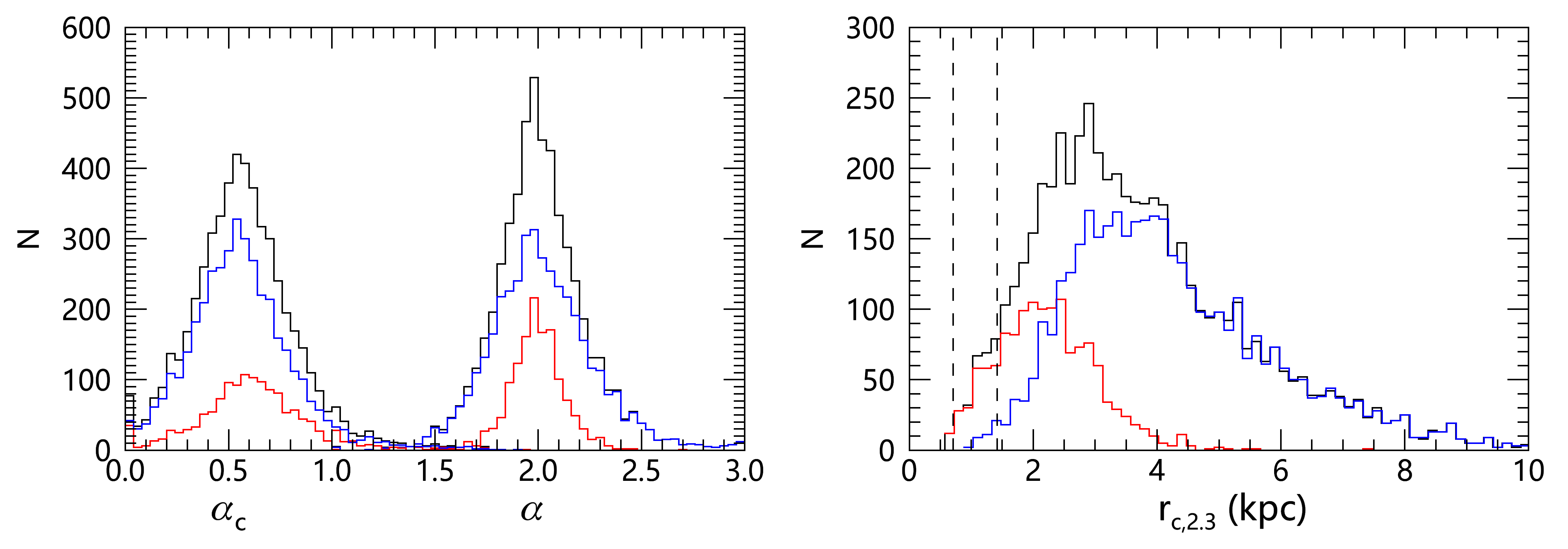}\\
  \caption{Left panel: distribution of inner slopes $\alpha_c$ and outer slopes $\alpha$ of mass density profile, where $\alpha_c$ are estimated from PL-S\'ersic profile fittings to the 3D light distribution with break radius $r_{c,2.3}$. Right panel: distribution of $r_{c,2.3}$. The two vertical dotted lines in the right panel mark the softening lengths in Illustris simulation for the stellar ($0.5\hkpc$) and dark matter particles ($1\hkpc$), respectively. In both panels, the red and blue histograms are for the elliptical and disk galaxies, respectively. The black histograms are for all of the galaxies.}\label{Fig:rc23} 
\end{figure*}

In the following, we show an example about the 3D fittings to the volume density profile of $\sim5000$ Illustris galaxies. Specifically, for each Illustris galaxy, we first fit its 3D light distribution within $90\%$ light radius using the PL-S\'ersic profile to get the break radius $r_c=r_{c,2.3}$ and the inner slope $\alpha_c=\alpha_{c,2.3}$. With fixed $r_c$ and $\alpha_c$ given by the light profile fitting, the volume mass density profile is then fitted by the BPL model with the same radii as those used for the light profile fitting. In Figure \ref{Fig:rc23}, we present the distributions of slopes (inner and outer slopes) in the left panel and break radii in the right panel for Illustris galaxies. We find that, with the definition of $r_{c,2.3}$, the inner part of the mass distribution can be clearly distinguished from the outer part. The outer slopes $\alpha$ are around $2$ and almost independent of galaxy type. The scatters of $\alpha$ are $\sim0.13$ for elliptical galaxies and $\sim0.23$ for disk galaxies. These results are in agreement with those found by the 2D ``joint'' fittings with FOV $6''\times6''$ shown in Figure 6 of \citet{2020ApJ...892...62D}.

In addition to the priors mentioned above, for the BPL model, we also examine the necessity of limiting the outer slopes (see Table \ref{tab:table}). We will find in the next section that the systematic errors in lens mass measurement outside Einstein radius can be largely reduced by confining the outer slope $\alpha$ to around 2 for the BPL model. For the other parameters in model fittings, uniform priors are applied with sufficiently large ranges.

Note that, in this paper, we single out individual Illustris galaxies for our analyses. Therefore, no large-scale structures are included, and consequently negligible effects from external convergence and shear. Even if a shear component exists due to correlated substructures or angular structures, its effect can be mimicked to some extent by ellipticity because of the well-known degeneracy between ellipticity and shear \citep{1993ApJ...417..450K,2021ApJ...915....4L}.

Also note that, for SIE and SPL models, we do not account for the central black holes and the above-mentioned limitations on axis ratios. For the BPL model, there are two cases. One is denoted as ``BPL-rigid'' with the above-mentioned priors for black hole mass ($m_b$), axis ratio ($q$), break radius ($r_c$), inner ($\alpha_c$), and outer ($\alpha$) slopes. The other is denoted as ``BPL-free'' for comparison, which retains the prior on the axis ratio, but does not have strong priors on the black hole mass, break radius, or slopes. Table \ref{tab:table} shows the specific priors on the radial parameters of the BPL model, where the SIE and SPL models are two of the special cases. We point out that there are two degrees of freedom in the radial direction for BPL-rigid and SPL models, while one for SIE and five for BPL-free models.

With the addition of axis ratios, position angles and two position parameters for a center, there are in total seven free parameters for a background source, and five, six, nine, and six free parameters for the SIE, SPL, BPL-free, and BPL-rigid lens mass models, respectively. We find that, for the SIE and SPL models, the MCMC fitting usually takes $\sim8$ CPU hours to reconstruct an SL image, while it takes about 24 and 20 CPU hours for the BPL-free and BPL-rigid models, respectively.

\begin{deluxetable}{ccccc}
\tablecaption{Priors on the radial parameters. \label{tab:table}}
\tablehead{
\colhead{Model} & \colhead{$\alpha$} & \colhead{$\alpha_c$} & \colhead{$r_c$} & \colhead{$m_b$}}
\startdata
SIE  & $2$ & $0$ & $0$ & $0$\\
SPL  & $1.2\sim2.8$ & $0$ & $0$ & $0$\\
BPL-free & $1.2\sim2.8$ & $0\sim2.8$ & $0\sim1\farcs5$ & $0\sim 10m_b^\star$\\ 
BPL-rigid & $1.8\sim2.2$ & $\alpha_{c,2.3}$ & $r_{c,2.3}$ & $m_b^\star$\\ 
\enddata
\tablecomments{The parameter $b$, which is not listed above, is set to be $0\farcs1\sim3\farcs5$ for all of the models. The symbol $m_b^\star$ denotes the black hole mass inferred from the total stellar mass of a lens galaxy.} 
\end{deluxetable}

\section{Results}\label{sec:result}
In this section, we investigate the performance of the BPL model as well as its special cases (\eg, the SIE and SPL models) in reconstructing the lensed images, lens mass, and source light distributions. We also examine the accuracy of estimating AL-weighted LOSVDs.

\subsection{Lensed image and lens mass reconstructions}
\subsubsection{Individual examples}
\begin{figure*}
  \centering
  \includegraphics[width=0.9\textwidth]{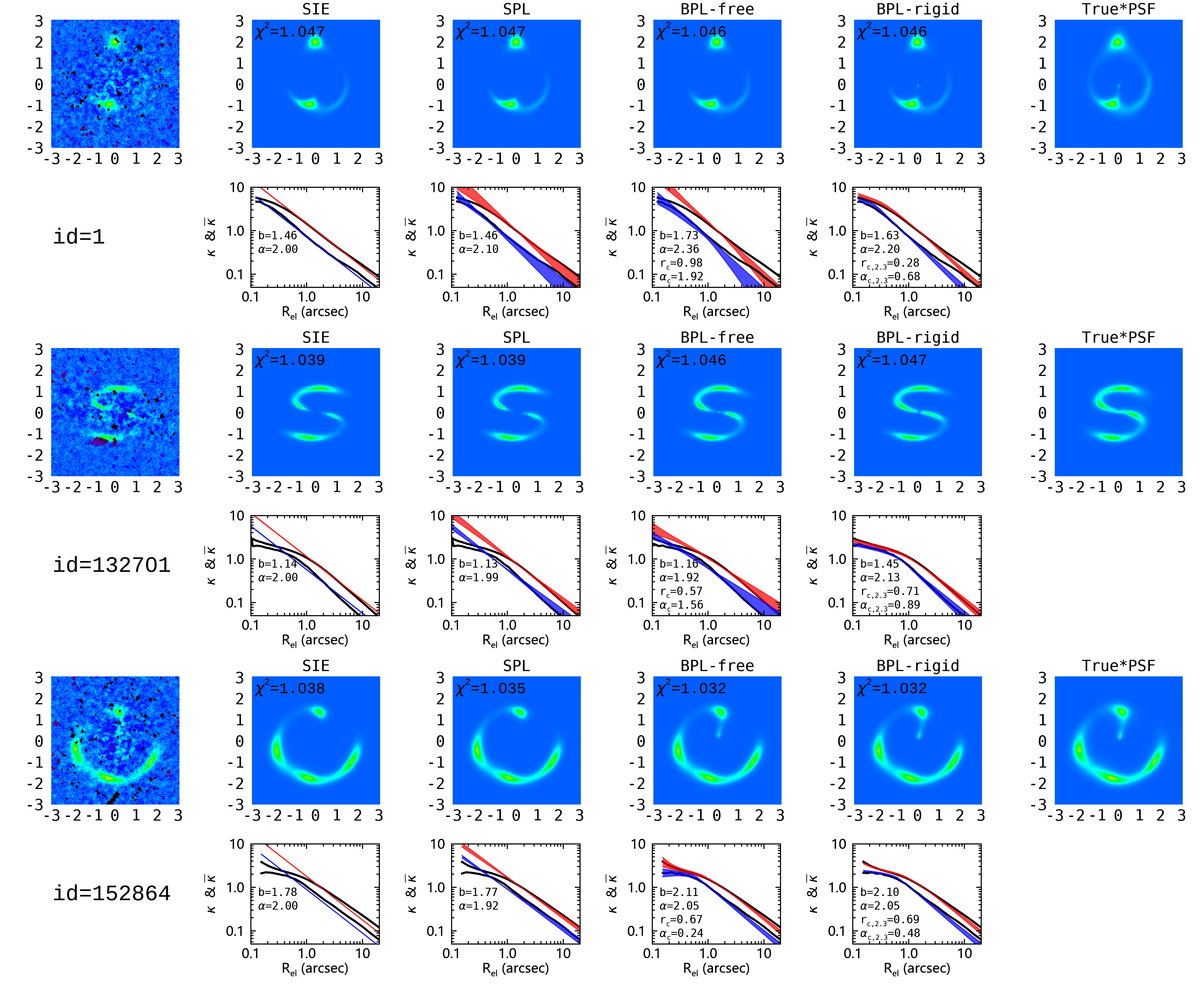}\\
  \caption{Lensed image and lens mass reconstructions for three single-exposure mock SL systems with ${\rm id}=1$ (top two rows), $132701$ (middle two rows), and $152864$ (bottom two rows). The leftmost and rightmost columns present the extracted and true lensed images in an FOV of $6''\times6''$, respectively. The middle four columns from left to right display the reconstructed images (odd rows), along with the corresponding reconstructed convergence profiles (even rows), based on SIE, SPL, BPL-free, and BPL-rigid models, respectively. The reduced $\chi^2$ values for the reconstructed images are shown at the left-top corner of the corresponding panels. In the panels about convergence profiles, the true radial convergence ($\kappa$) and mean convergence ($\bar{\kappa}$) profiles are shown by black lines, where $\kappa(R_{\rm el})<\bar{\kappa}(R_{\rm el})$ and $R_{\rm el}$ is the elliptical radius. The blue and red shaded bands indicate the $\pm3\sigma$ error bands around the best-fit profiles. Also shown are the best-fit values relevant to the radial profiles.}\label{Fig:img_recon}
\end{figure*}

As an example, Figure \ref{Fig:img_recon} shows the reconstruction of lensed images and radial convergence profiles of three mock SL systems with single exposure. By comparing with the true lensed images shown in the last column and the true convergence profiles shown as the black lines, one can notice from the second and third columns that the SIE and SPL models can reproduce the lensed images around Einstein radius quite well, but cannot recover the lensing features toward the center and the inner density profile accurately. The fourth column of Figure \ref{Fig:img_recon} demonstrates that, the BPL-free model, which is more flexible in the radial direction, succeeds in recovering the central lensing features as well as the inner convergence profile for the lens $152864$ (the number here is the subhalo id in the Illustris-1 simulation), but fails for the other two lenses with ${\rm id}=1$ and $132701$.

We show the results of BPL-rigid model fittings in the fifth column of Figure \ref{Fig:img_recon}. It is amazing that, for all the three SL systems, the lensed images and radial convergence profiles can be recovered with high fidelity, although the relevant $\chi^2$ values may not be the smallest. These results illustrate the necessity of adding priors to the radial density profile, especially in the inner region where the central images, if they exist, may be too faint to be detected due to the significant contamination of lens light distribution. The BPL model has the ability to determine the possible central images by using the lens light distribution to constrain the inner density profile.

We know that the image quality can be improved by multiple exposures. However, by inspecting the lensed images extracted from the SL images with four exposures, we find the contamination from lens light cannot be effectively reduced by increasing exposure time for galaxy-scale lenses. The extracted lensing features for multiple exposures are almost the same as those for single exposure, and so are the reconstructed lensed images and convergence profiles, although the lensing images with more exposures suffer less from the cosmic rays and Poisson noise.

\begin{figure*}
  \centering
  \includegraphics[width=0.9\textwidth]{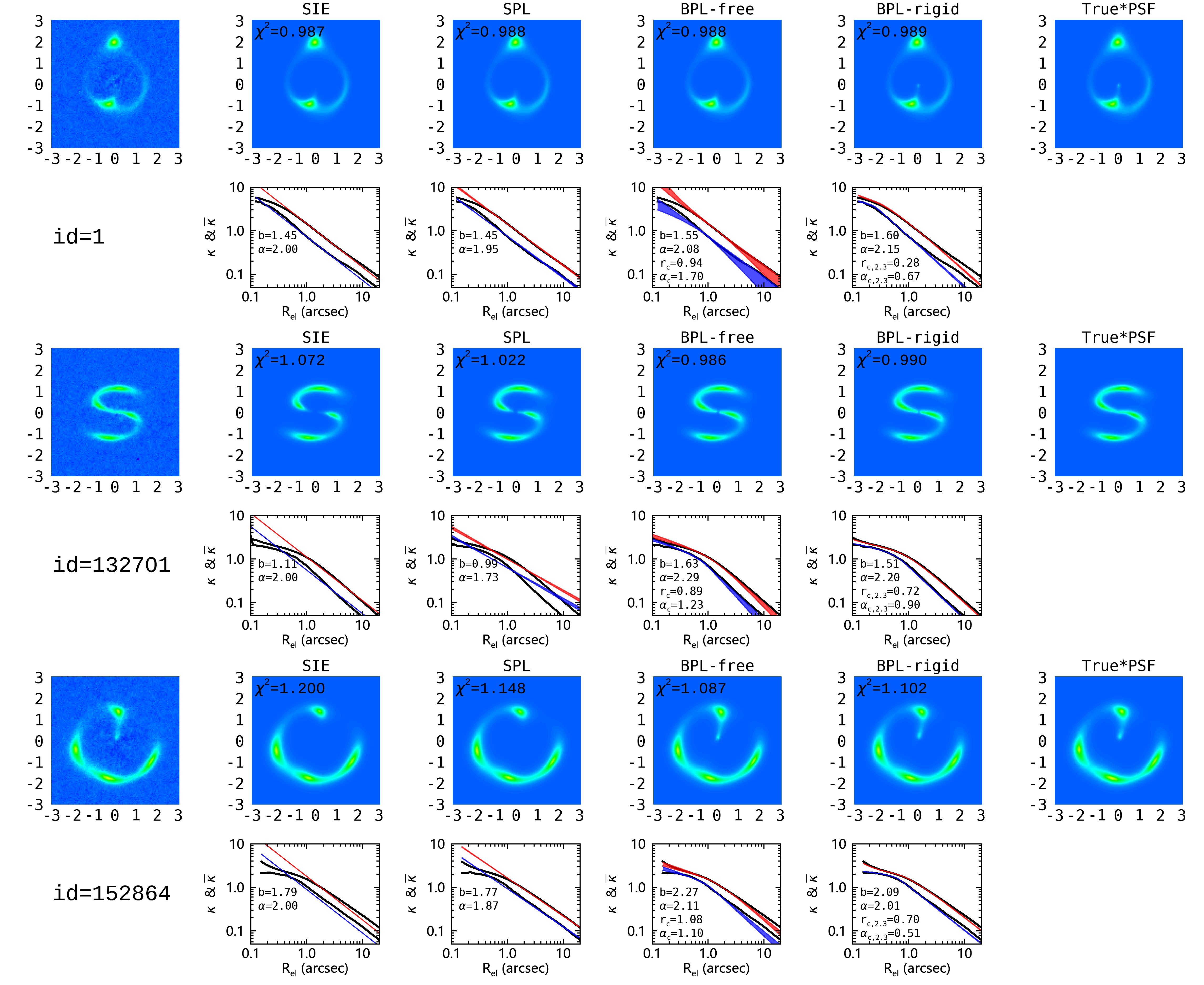}\\
  \caption{Similar to Figure \ref{Fig:img_recon} but for multiple exposures with perfect subtraction of foreground lens light distribution.}\label{Fig:img_recon0}
\end{figure*}

An ideal case is that the foreground lens light is perfectly subtracted from SL images. We investigate this case based on the mock SL images with four exposures (2200s in total). Similar to Figure \ref{Fig:img_recon}, Figure \ref{Fig:img_recon0} shows the relevant results, where the first column displays the lensed images left after subtracting the true lens light directly. Compared to Figure \ref{Fig:img_recon}, the extracted lensed images become clearer, although there are still Poisson noises from lens light.

Figure \ref{Fig:img_recon0} shows that, even if there are no lens light contaminations, it is impossible for SIE and SPL models to recover the lensed images and convergence profiles in the central region. The BPL-free model can now reconstruct well the lensed images and convergence profiles of the lens $132701$, but still fails for lens 1, whose central image disappears in noise. When looking at the accuracy of mass measurements rather than $\chi^2$ values for image reconstructions, the BPL-rigid model with strong radial priors still performs best.

Both in Figures \ref{Fig:img_recon} and \ref{Fig:img_recon0}, the blue (red) shaded bands show the $\pm3\sigma$ error ranges of the reconstructed radial (mean) convergence profiles. We notice that, compared to their deviations from true convergence profiles, the uncertainties estimated from model fittings are subdominant in most cases. We thus do not pay attention to the precision or statistical errors of the relevant quantities for each lens system, but focus on the statistics about the best-fit values or reconstructions.

\subsubsection{Einstein radii}
\begin{figure*}
  \centering
  \includegraphics[width=0.9\textwidth]{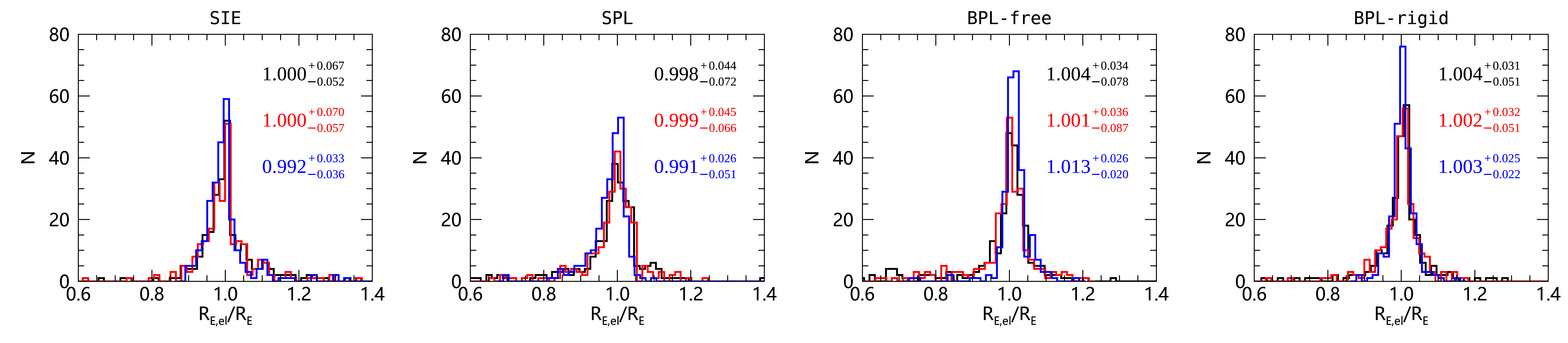}\\
  \caption{Distributions of the ratios between best-fit elliptical Einstein radii $R_{\rm E,el}$ and their true values $R_{\rm E}$. In each panel, the black, red, and blue histograms are for the cases with single exposure, four exposures, and four exposures with perfect subtraction of lens light, respectively. For each histogram, also shown is its median value along with the $68\%$ confidence interval.}\label{Fig:rein}
\end{figure*}

For an elliptical lens, its Einstein radius inferred from a model fitting is defined as the elliptical radius $R_{\rm E,el}$ within which the mean convergence satisfies $\bar{\kappa}=1$. Figure \ref{Fig:rein} displays the histogram distribution of the ratio of the best-fit Einstein radius $R_{\rm E,el}$ and the directly measured circular Einstein radius $R_{\rm E}$, for different model fittings and noise levels. We find that, independent of mass models and noise levels, the Einstein radius $R_{\rm E}$ can be approximated by $R_{\rm E,el}$ with controllable scatter of biases, where the median value of the biases (termed ``median bias'' hereafter) is typically subpercent. This indicates that $R_{\rm E,el}$ is a good representative of $R_{\rm E}$ for a lens with complex structures, although a little bias may exist between them for a smooth elliptical mass distribution.

Specifically, in each panel of Figure \ref{Fig:rein}, the red histograms for the multiple exposures are very similar to the black histograms for the single exposure. Their scatters are almost the same, indicating the limitation of using more exposures to get a better inference of Einstein radius. This is mainly because the light contamination from the main lens itself cannot be effectively reduced by more exposures. If we artificially subtract the true lens light, as shown by the blue histograms and its $68\%$ central confidence interval, the accuracy in $R_{\rm E}$ estimation can be largely improved, depending on the adopted lens mass models.

Also focusing on the black and red histograms, we find that the scatters of the ratio $R_{\rm E,el}/R_{\rm E}$ for SIE, SPL, and BPL-free models are typically greater than $5\%$ and can even reach up to $\sim9\%$. For the BPL model with inner density profile predetermined, the scatter of $R_{\rm E,el}/R_{\rm E}$ can be reduced to $\sim5\%$. In any case, we notice that the scatter of $R_{\rm E}$ estimation is unlikely to be less than $2\%$, even for the idealized case without lens light contamination as indicated by the blue histograms in Figure \ref{Fig:rein}.

\subsubsection{Radial convergence profiles}\label{subsec:rkappa}
\begin{figure*}
  \centering
  \includegraphics[width=0.9\textwidth]{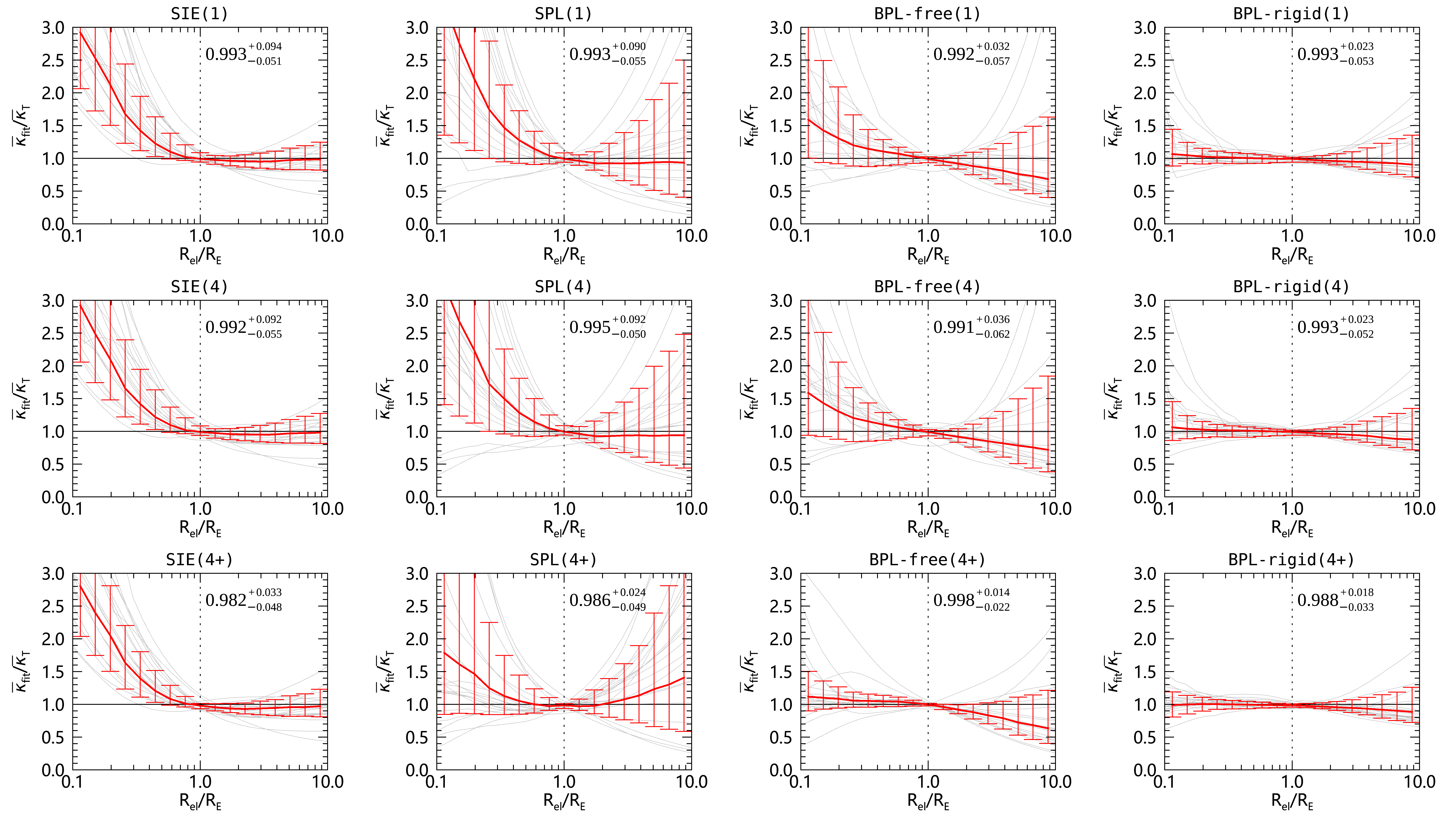}\\
  \caption{Comparison of reconstructed mean convergence profiles $\bar{\kappa}_{\rm fit}$ and true ones $\bar{\kappa}_{\rm T}$ as a function of elliptical radius $R_{\rm el}$ scaled by true Einstein radius $R_{\rm E}$. The lens mass models used for image reconstructions are indicated by the titles, with $(1)$, $(4)$, and $(4+)$ signifying the single exposure, four exposures, and four exposures with perfect subtraction of lens light, respectively. We show 20 gray lines in each panel, which correspond to 20 randomly selected SL systems. The red lines show the median trend of $\bar{\kappa}_{\rm fit}/\bar{\kappa}_{\rm T}$ for the 257 grade-A lenses. The error bars show the $68\%$ central confidence intervals at the corresponding radii. The vertical dotted line in each panel marks the radius where $R_{\rm el}=R_{\rm E}$ (\ie, the true Einstein radius), at which the median and $68\%$ confidence interval of the biases on the mean convergence profile are reported.}\label{Fig:mkappa}
\end{figure*}

We now inspect the accuracy of reconstructed radial convergence profiles over a wide range of radii from $0.1R_{\rm E}$ to $10R_{\rm E}$. Figure \ref{Fig:mkappa} shows the statistical results of mean convergence profiles as a function of elliptical radius. Note that, close to Einstein radius, both the median value and the scatter of the biases approach the minimum regardless of lens mass models or noise levels. Away from Einstein radius, the scatters tend to be larger and larger, indicating the difficulty in recovering the mass distribution in these regions.

It is also evident that the median biases for the SIE and SPL models are rather large within Einstein radius, which can be greatly reduced by using the more flexible BPL model. Especially for the BPL model with rigid priors, the median biases are typically less than $5\%$ within $R_{\rm E}$, and less than $2\%$ at $R_{\rm E}$. For the BPL-free model, as illustrated by the third column of Figure \ref{Fig:mkappa}, only when the lens light distribution is perfectly subtracted can the mass distribution be well recovered within $R_{\rm E}$ due to the larger signal-to-noise ratio of central lensing features.

Outside Einstein radius, an amazing result is that the median biases for the SIE model are typically less than $10\%$, and the scatters on the biases are also not too significant. The SPL model still produces large scatters away from Einstein radius. We notice that the BPL-free model is not good at constraining the convergence profiles at radii larger than $R_{\rm E}$. By adding strong priors to the radial parameters, the BPL-rigid model can recover the mean convergence profiles within $3R_{\rm E}$ quite well, with median biases less than $5\%$ and controllable scatter.

It is worth noting that the large biases and scatters within $R_{\rm E}$ for SIE and SPL models may not be true for real galaxy lenses, which lack large and flat cores. Nonetheless, the first two columns of Figure \ref{Fig:mkappa} demonstrate that, compared with the BPL-rigid model, the SIE and SPL models can result in larger biases in reconstructing more complex lens mass distribution.

\begin{figure*}
  \centering
  \includegraphics[width=0.9\textwidth]{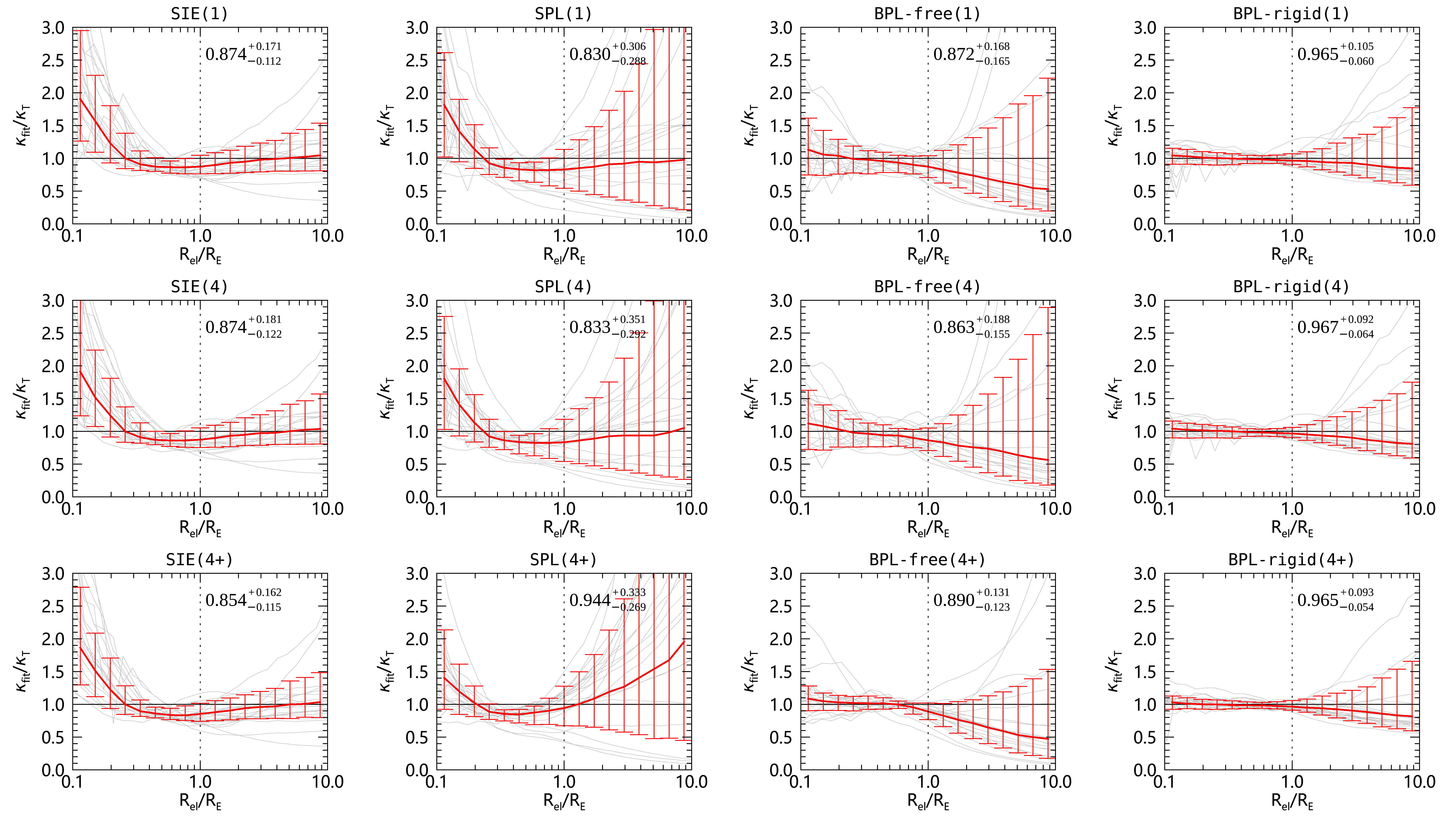}\\
  \caption{Similar to Figure \ref{Fig:mkappa}, but for the radial convergence profiles $\kappa(R_{\rm el})$.}\label{Fig:kappa}
\end{figure*}

In addition to the mean convergence profiles, we are also interested in the radial convergence profiles, especially the convergence $\kappa_{\rm E}$ at Einstein radius. In Figure \ref{Fig:kappa}, we exhibit the comparisons of the reconstructed convergence profiles from the true profiles. We realize again that, without reasonable priors or additional information on the mass distribution, it is impossible to constrain the radial convergence profiles over a wide range of radii based solely on lensed images. As shown in the last column of Figure \ref{Fig:kappa}, the BPL-rigid model outperforms all the other models. Within $R_{\rm E}$ and $3R_{\rm E}$, the median biases are typically no more than $5\%$ and $10\%$, respectively.

Figure \ref{Fig:kappa} also demonstrates that the scatters on the biases in $\kappa_{\rm E}$ measurements are quite large, especially for the SPL model fittings. Large systematic errors in $\kappa_{\rm E}$ make accurate measurements of $H_0$ from time-delay cosmography difficult, because of the well-known scaling relation $H_0\propto1-\kappa_{\rm E}$ \citep{2020MNRAS.493.1725K,2021MNRAS.501.5021K}. The fractional error in $H_0$ can be roughly evaluated by $f_{H_0}=(1-\kappa_{E,\rm fit})/(1-\kappa_{E,\rm T})-1$, where $\kappa_{E,\rm fit}$ and $\kappa_{E,\rm T}$ are the fitted and true convergences at $R_{\rm E}$, respectively. By calculating $f_{H_0}$, we find its median reaches $\sim20\%$ for SIE, SPL, and BPL-free models, with much larger scatters. For the BPL-rigid model, the median bias in $H_0$ is less than $6\%$. Also taking into account the large scatters of $f_{H_0}$ and impacts other than lens modeling, we conclude that, even for the BPL-rigid model, $H_0$ can probably only be constrained to an accuracy of $\sim10\%$ for a single SL time-delay system.

\subsection{Background source light reconstructions}
\begin{figure*}
  \centering
  \includegraphics[width=0.7\textwidth]{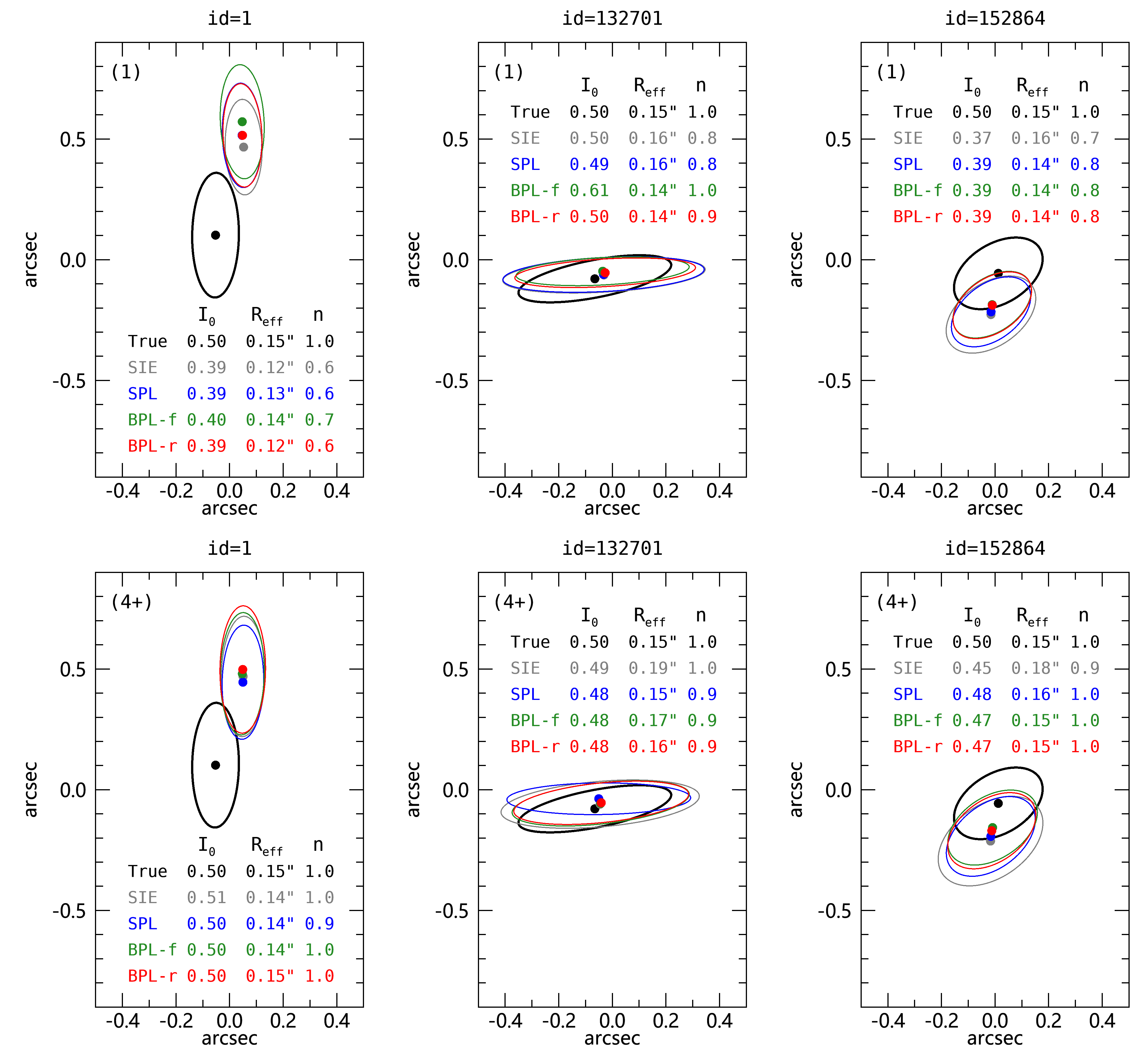}\\
  \caption{Source light reconstructions for three SL systems from left to right with ${\rm id}=1$, $132701$, and $152864$, respectively. The top panels show the results for a single exposure, while the bottom panels show the results for multiple exposures without lens light contamination. In each panel, the ellipses show the isophotes at effective radii of the corresponding source light distribution (black for the true and colored for the reconstructed), and dots mark their centers. Also displayed are the S\'ersic profile values of $I_0$, $R_{\rm eff}$ and $n$ corresponding to the ellipses.}\label{Fig:spt}
\end{figure*}

In Figure \ref{Fig:spt} we show the source light reconstructions for the three SL systems with ${\rm id}=1$, $132701$, and $152864$, whose mass reconstructions are shown in Figures \ref{Fig:img_recon} and \ref{Fig:img_recon0} as examples. The top panels display the source light properties estimated in the case of single exposure, while the bottom panels are for the multiple exposures with perfect removal of lens light. By looking at the ellipses, we notice that the centers, ellipticities, and position angles of the reconstructed source light distribution are highly correlated between different lens mass models, although large offsets may exist compared to the true values. Statistically, we find that the correlation coefficients for centers, ellipticities, and position angles predicted by different lens mass models are all greater than $0.8$. There are also weak correlations for parameters $I_0$ and $n$ between different lens model fittings to the same lensed images, as shown by the parameter values listed in each panel.

By comparing the corresponding ellipses in the bottom panels to those in the top panels, we realize that the improvement in image quality cannot effectively reduce the center offsets of source galaxies. However, if there is no visible contamination from lens light, the amplitude $I_0$ and S\'ersic index $n$ become more consistent with the input values.

\begin{figure*}
  \centering
  \includegraphics[width=0.9\textwidth]{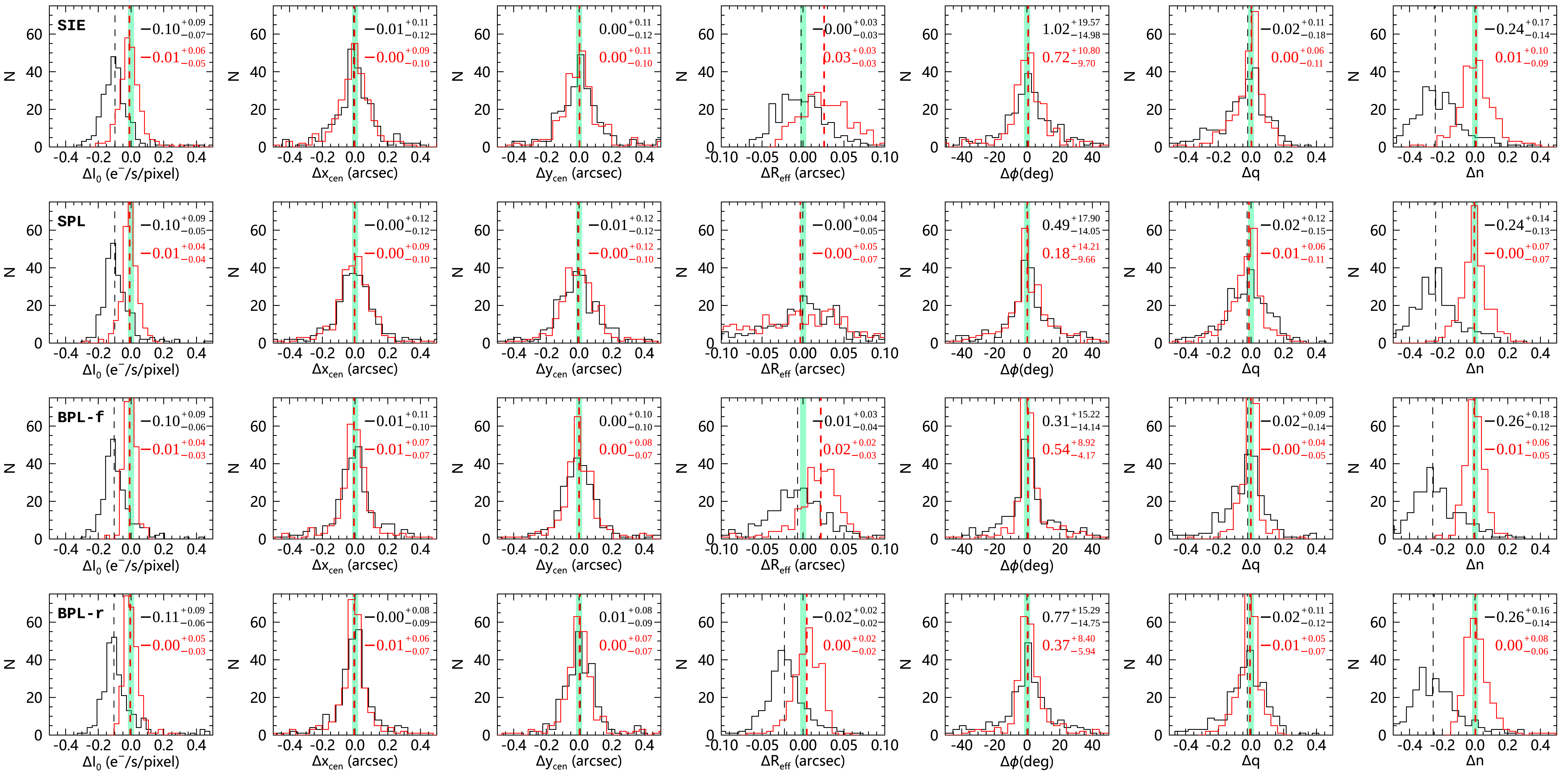}\\
  \caption{Histograms of the differences between estimated and true S\'ersic profile parameter values for the source light distribution. From top to bottom, the panels show SIE, SPL, BPL-free, and BPL-rigid models, respectively. From left to right, the relevant parameters are, respectively, the central light intensity $I_0$, source center $(x_{\rm cen},y_{\rm cen})$, effective radius $R_{\rm eff}$, position angle $\phi$, axis ratio $q$ and S\'ersic index $n$. The black histograms represent the cases for single exposure, while the red histograms are for multiple exposures without lens light contamination. The numbers in each panel denote the medians along with central $68\%$ confidence intervals of the corresponding histograms, where the medians are marked by vertical dashed lines. The vertical green bands indicate the positions of zero-deviation. Note that the true values for $I_0$, $R_{\rm eff}$, and $n$ are $0.5~\rm e^-s^{-1}pixel^{-1}$, $0\farcs15$, and $1.0$, respectively.}\label{Fig:srcpars}
\end{figure*}

Figure \ref{Fig:srcpars} exhibits the distributions of the differences between estimated and input source parameter values. The black histograms are for the single-exposure cases, while the red histograms are for multiple exposures without lens light contamination. We do not plot the histograms for multiple exposures with significant lens light contamination, which are very similar to the histograms for single-exposure cases. It is now more clear that, although the scatters on the deviations are significant, there are basically no median biases for the centers, position angles, and ellipticities of the source galaxies.

For single-exposure cases, the central light intensity $I_0$ and S\'ersic index $n$ tend to be smaller, demonstrating the significant effect of lens light contamination on the determination of source light density profile. The median biases for $I_0$ and $n$ disappear when the lens light is perfectly subtracted.

Of the measurements of source parameters, the most complicated one is the inference of the effective radius that is not only sensitive to the lens light contamination but also lens mass models. The lens light contamination may spoil the lensed images, while the lens mass models may introduce bias in mass distribution around Einstein radius and thus the magnifications.

We show the histograms of the deviations of the effective radii from their truth in the fourth column of Figure \ref{Fig:srcpars}. We notice that, for single-exposure cases, the SIE, SPL, and BPL-free models coincidentally result in median-unbiased estimates of the effective radius but with broad distributions. However, for the cases without lens light contamination, the effective radii derived from SIE and BPL-free models tend to be larger, as indicated by the red histograms. The median bias of $R_{\rm eff}$ estimation for the SPL model remains almost unchanged since the reconstructed mass distribution for the SPL model instead has smaller biases around Einstein radii (see the bottom panel in the second column of Figure \ref{Fig:kappa}). By contrast, when there is no lens light contamination, the BPL-rigid model can provide a more reliable and accurate estimate of $R_{\rm eff}$ as well as other source galaxy parameters.

We now pay attention to the large error in source center determinations. As indicated by Figures \ref{Fig:spt} and \ref{Fig:srcpars}, there are cases with large displacements of source centers. For example, the ${\rm id}=1$ with single exposure shows a source center displacement of $\sim 0\farcs41$. We look for the cause of large offsets of source centers by comparing the true and reconstructed lens mass distribution out to the outskirts of lens halos, and also the deflection fields. As expected, we find that many lenses show complexity of mass distribution in the inner or outer regions, which may induce additional deflections around Einstein radii compared to the estimated elliptical mass distribution.

In this paper, we define additional deflections as the difference between the true and reconstructed deflections. We approximate the additional deflections with their average within the Einstein radius, denoted as a constant vector $(\Delta\alpha_{x,0},\Delta\alpha_{y,0})$. We find the standard error of the additional angles within Einstein radius is less than $0\farcs05$ for $60\%$ cases and $0\farcs09$ for $90\%$ cases. In view of the fact that there are hundreds to thousands of pixels within Einstein radius, the error of the average is negligible. It should be mentioned that the average of deflections within Einstein radius is zero for an elliptically symmetric mass distribution or a mass sheet. A nonzero value of the average implies the asymmetry of the lens mass distribution.

\begin{figure*}
  \centering
  \includegraphics[width=0.9\textwidth]{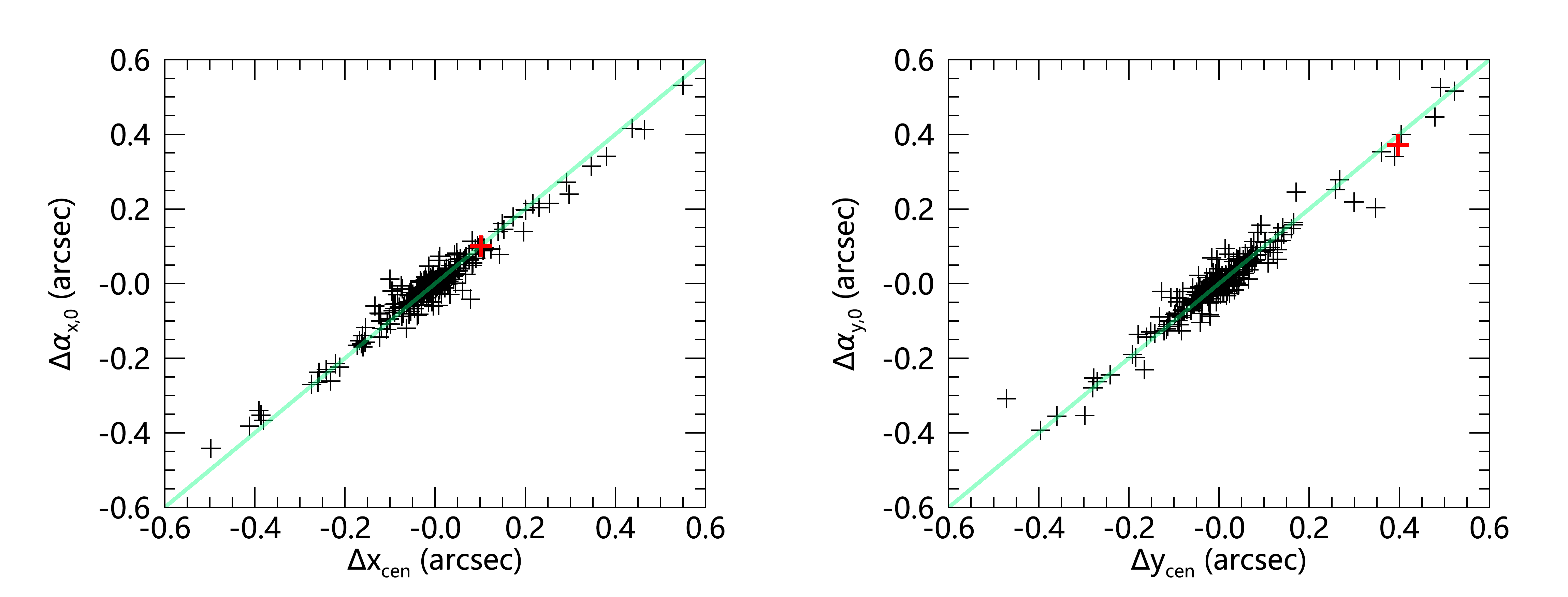}\\
  \caption{Comparisons between the estimated constant deflections $(\Delta\alpha_{x,0},\Delta\alpha_{y,0})$ and source center offsets $(\Delta x_{\rm cen},\Delta y_{\rm cen})$ for the BPL-rigid modeling on lensed images without lens light contamination. The left and right panels are for the $x$- and $y$-directions, respectively. The red pluses denote the case with ${\rm id}=1$.}\label{Fig:scen} 
\end{figure*}

Figure \ref{Fig:scen} shows the strong correlation between $(\Delta\alpha_{x,0},\Delta\alpha_{y,0})$ and $(\Delta x_{\rm cen},\Delta y_{\rm cen})$ for the BPL-rigid modeling on lensed images without lens light contamination. The results show that the large offsets of source galaxy centers can be largely explained by constant deflection angles around the center of lenses. For instance, for the case ${\rm id}=1$, the amplitude of its additional constant deflection is $\sim 0\farcs38$, in line with its source center offset of $\sim 0\farcs41$. These results demonstrate the large offset of source centroid is mainly caused by the asymmetric mass distribution rather than a mass-sheet transform, which has no contribution to the average of deflection angles within Einstein radius.

\subsection{AL-weighted LOSVDs}
AL-weighted LOSVDs, denoted as $\sigma_\parallel={\langle\sigma_\parallel^2\rangle}^{\frac{1}{2}}$ hereafter, can provide complementary constraints on the lens mass distribution, and may help us break MSD. However, in order to implement this idea, we need to ensure there is no bias from modeling itself. \citet{2020ApJ...892...62D} inspected the capability of the BPL model on estimating the AL-weighted LOSVDs based on the spherical Jeans modeling (see Section \ref{subsec:losvd}), where the mass and light distribution for each galaxy are directly fitted by the BPL and PL-S\'ersic profiles, respectively, with the same inner slope and break radius. It was found that a systematic offset exists between the predicted AL-weighted LOSVDs and their true values. This velocity dispersion bias, mainly due to the projection effect, can be systematically corrected by $b_\sigma\simeq1.015q_\star^{-0.07}$ with $q_\star$ being the axis ratio of the lens light distribution.

We in this subsection look at the consistency between the predicted and true AL-weighted LOSVDs from simulations in the context of SL observations, where the former is evaluated based on the reconstructed lens mass and directly fitted lens light density profile. For SIE and SPL lens mass models, the S\'ersic profile is adopted for lens light fittings, whereas for BPL models, the PL-S\'ersic profile is adopted. More specifically, for BPL-free modeling, the inner density profile of lens light distribution is determined by the lens mass reconstruction. However, for BPL-rigid modeling, the inner slope and break radius of the lens mass distribution are fixed to be $\alpha_{c,2.3}$ and $r_{c,2.3}$, respectively, and are determined by fitting to the lens light distribution. Also note that, in the estimation of AL-weighted LOSVD for each lens, the global velocity anisotropy is adopted and assumed to be known.

\begin{figure*}
  \centering
  \includegraphics[width=0.9\textwidth]{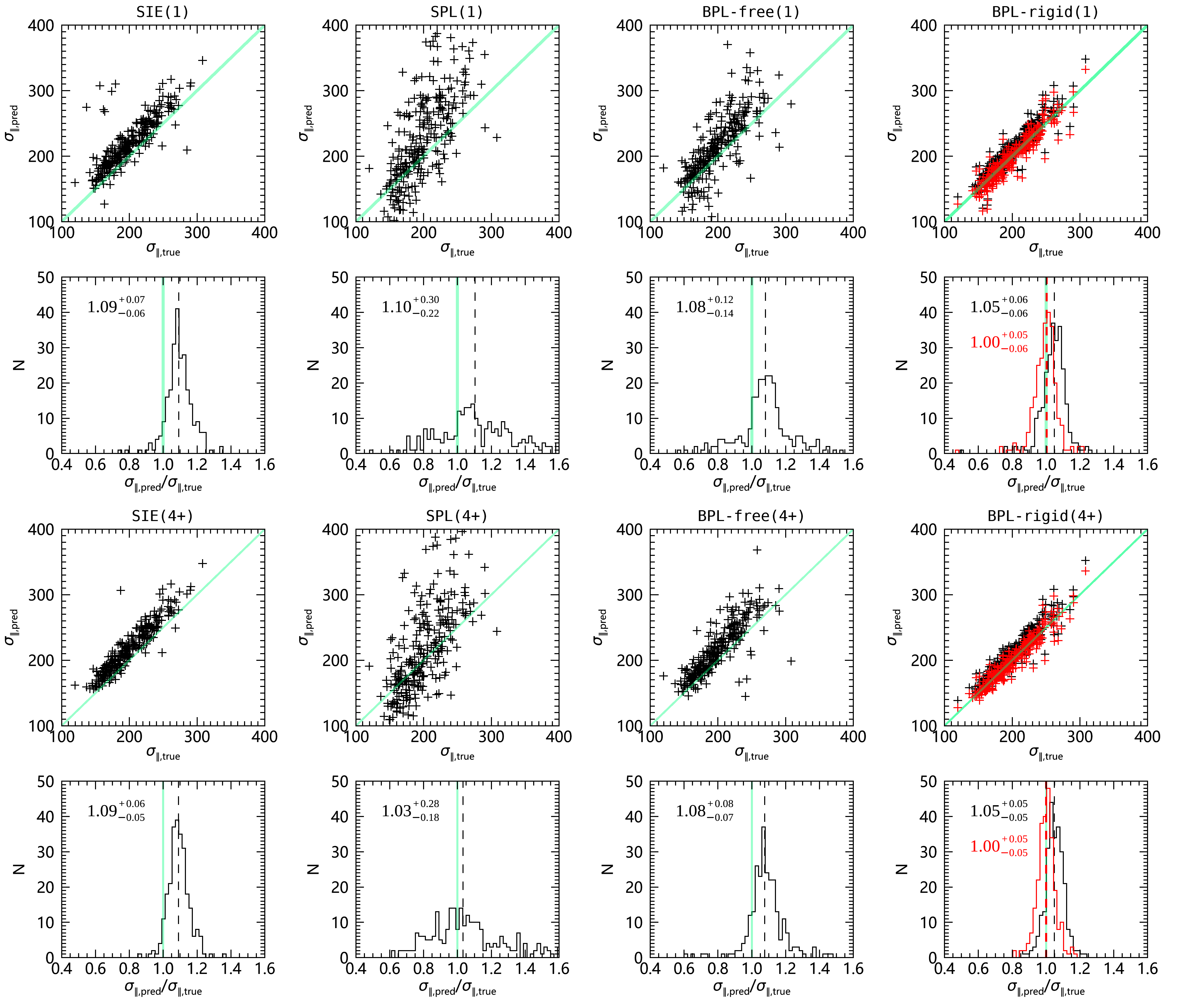}\\
  \caption{Comparisons of the predicted AL-weighted LOSVDs $\sigma_{\rm\parallel,pred}$ to their true values $\sigma_{\rm\parallel,true}$. From left to right, the columns correspond to SIE, SPL, BPL-free, and BPL-rigid models, respectively. The top two rows show the results for the single exposure cases, while the bottom two rows are for multiple exposures without lens light contamination. The scatter diagrams present the one-to-one comparisons, with green lines indicting the identity lines. The histograms show the corresponding distributions of ratio $\sigma_{\rm\parallel,pred}/\sigma_{\rm\parallel,true}$, where vertical dashed lines show the medians of the distributions and green lines mark the position without bias. The numbers represent the medians of the corresponding histograms along with $68\%$ confidence intervals. In the rightmost panels, the red pluses and histograms correspond to the results by accounting for the velocity dispersion bias $b_\sigma$ for BPL models.}\label{Fig:allosvd}
\end{figure*}

Figure \ref{Fig:allosvd} displays the comparisons of the predicted and true AL-weighted LOSVDs, where the top two rows correspond to the single exposure cases, and the bottom two rows are for the four exposures with perfect subtraction of lens light. We do not present the results for multiple exposure cases with lens light contamination, which are very similar to those for single exposure cases. We can see from the black histograms or the black scatter diagrams that there are positive biases in $\sigma_\parallel$ estimations for most cases. The median biases are typically larger than $5\%$ and have a certain dependence on the lens mass models. We notice that the scatter on the biases is more vulnerable to lens mass models. For instance, the scatter for the SPL model can reach up to $30\%$ while it is only about $5\%$ for the BPL-rigid model.

By comparing the bottom two rows with the top two rows, we also note that the improvement in the quality of lensed images has little effect on the median biases of $\sigma_\parallel$ predictions but may reduce the scatter of the biases significantly for some lens mass models. For example, the scatter for the BPL-free model can be reduced from $14\%$ to $8\%$. In contrast, the distributions of the fractional deviations for the BPL-rigid model are quite stable, almost insensitive to lens light contamination. After correcting for the projection effect in the BPL-rigid model fittings, \ie, dividing the predicted $\sigma_\parallel$ by $b_\sigma$ for each lens, the red scatter diagrams and histograms show that the finally obtained $\sigma_\parallel$ can match its true value to within $6\%$. We note that, by comparing the true and predicted velocity dispersions, an effective external convergence can be estimated according to $\kappa_{\rm ext}=1-\sigma_{\rm\parallel,true}^2/\sigma_{\rm\parallel,pred}^2$. We thus expect that, for an SL system with velocity dispersion measured from spectroscopic observations, it is possible to infer its external convergence to within $12\%$ accuracy using the BPL-rigid model.

\subsection{Model evaluation}
In the previous subsections, we examined the accuracy of the reconstructed or predicted quantities by comparing them with their true values directly. We now look at the commonly adopted criteria, \eg, the reduced $\chi^2$, Akaike information criterion (AIC) and Bayesian information criterion (BIC), to evaluate the quality of model fits \citep{1995JASA...90..773,2020A&A...639A.101M,2021MNRAS.501.5021K}. The AIC and BIC are estimated by ${\rm AIC}=2k+\chi^2$ and ${\rm BIC}=k\ln n+\chi^2$, respectively, where $k$ is the number of free parameters, $\chi^2$ is defined by Equation (\ref{eq:chi2scale}), and $n$ is the number of pixels with nonzero weights. 

In Table \ref{tab:table2}, we summarize the median values of the reduced $\chi^2$, $\Delta$AIC, and $\Delta$BIC for three noise levels and five lens models. We note that, for cases (1) and (4) that suffer from nonnegligible lens light contamination, the BPL-free model with more degrees of freedom leads to the minimum values of the estimators, indicating its good performance on the reconstruction of extracted lensed images. However, the supposed ``True'' model, which recovers the true lens mass distribution and true lensed images perfectly, gives the maximum values of the criteria, indicating there are large deviations of the extracted lensed images from the true ones due to the lens light contamination. The BPL-rigid model, which can recover the lens mass distribution more accurately than SIE, SPL, and BPL-free models as indicted by Figures \ref{Fig:mkappa} and \ref{Fig:kappa}, also performs poorly.

According to the reported values of case (4+), when lens light is perfectly subtracted, the ``True'' model becomes the best and the BPL models are preferred over the SIE and SPL models. We find that the BPL-rigid model is still not as good as the BPL-free model in recovering the lensed images, although the former can reconstruct the mass distribution over a wider range of radii and make more reliable predictions of AL-weighted LOSVDs. These results demonstrate that, due to lensing degeneracies and lens light contamination, the estimators for model selection investigated here may be good at distinguishing lens models in reproducing the lensed images, but may not be suitable for evaluating lens mass models in measuring the lens mass distribution.

\begin{deluxetable*}{ccccccc}
\tablecaption{Model evaluations with the reduced $\chi^2$, AIC, and BIC values. \label{tab:table2}}
\tablehead{
\colhead{$ $} & \colhead{$ $} & \colhead{SIE}  & \colhead{SPL} & \colhead{BPL-free} & \colhead{BPL-rigid} & \colhead{True}}
\startdata
&$\chi^2$   & $1.055$  & $1.048$  & $1.046^\ast$  & $1.051$  & $1.078$\\
(1)&$\Delta$AIC   & $    (12678.7)$  & $  -36.6$  & $  -64.2^\ast$  & $  -19.5$  & $  204.3$\\
&$\Delta$BIC   & $    (12767.5)$  & $  -29.4$  & $  -35.2^\ast$  & $  -12.2$  & $  211.8$\\
\hline
&$\chi^2$   & $1.051$  & $1.035$  & $1.031^\ast$  & $1.042$  & $1.084$\\
(4)&$\Delta$AIC   & $    (13493.9)$  & $  -75.5$  & $ -131.6^\ast$  & $  -47.0$  & $  378.4$\\
&$\Delta$BIC   & $    (13583.8)$  & $  -67.9$  & $ -102.2^\ast$  & $  -39.5$  & $  386.0$\\
\hline
&$\chi^2$   & $1.072$  & $1.010$  & $0.990$  & $1.001$  & $0.966^\ast$\\
(4+)&$\Delta$AIC   & $    (13794.2)$  & $ -343.2$  & $ -829.0$  & $ -436.9$  & $-1230.5^\ast$\\
&$\Delta$BIC   & $    (13884.5)$  & $ -335.6$  & $ -799.7$  & $ -429.6$  & $-1222.9^\ast$\\
\enddata
\tablecomments{Shown are the median values of the reduced $\chi^2$, $\Delta$AIC, and $\Delta$BIC, where the $\Delta$ denotes the difference relative to the SIE model for which the median values of AIC and BIC are displayed instead. The ``True'' lens model, which recovers the true lens mass distribution and true lensed images perfectly, is assumed to have the same number of free parameters as the BPL-rigid model. The smaller the value, the better the model in reproducing the extracted lensed images. The minimum value for each criterion is marked with asterisk.}
\end{deluxetable*}
\section{Conclusion and discussions}%
The BPL model proposed by \citet{2020ApJ...892...62D} is a lens mass model with four free parameters in the radial direction. It is a more flexible mass model with analytical deflections. In this paper, we examine the performance of the BPL model on the lens mass and source light reconstructions, as well as the predictions of AL-weighted LOSVDs.

In order to quantify the systematics in the relevant quantities, we implement an end-to-end test by starting with the creation of mock SL observations with various noises, \eg, sky background, cosmic rays, PSF, and readout noise, where the lenses are selected from the simulated galaxies in the Illustris-1 simulation. To extract the lensed images, the B-spline technique is applied to fit the lens light distribution. We finally have $\sim260$ ``grade-A'' lenses left with two different exposure times (420~s for a single exposure and 2200~s for four exposures) for further analyses.

We use forward modeling to reproduce the lensed images, along with the lens mass and source light distributions. Four lens mass models, \ie, the SIE, SPL, BPL-free, and BPL-rigid models, are investigated, where the SIE and SPL models are in fact the special cases of the BPL model. The SIE model has a slope of 2, while the SPL model has a single free slope. The difference between the BPL-free and BPL-rigid models is whether or not strong priors are applied to the radial profile of the BPL model.

By looking at the lensed image reconstructions, we find that, if there are no obvious central images identified, the extracted lensing features around Einstein radii can always be recovered fairly well, almost independent of lens mass models. Based solely on the $\chi^2$ values of the image fittings, it is hard to judge which lens mass model performs best in reconstructing the lens mass distribution. If central images are evident in the extracted lensed images, the BPL models outperform the single power-law models in both lensed image and lens mass reconstructions. On the other hand, if central images are submerged in lens light, the BPL-rigid model is capable of determining the missing central images.

We investigate the accuracy of Einstein radius measurements by comparing the elliptical Einstein radii $R_{\rm E,el}$ inferred from model fittings with the ``true'' circular Einstein radii $R_{\rm E}$. Although different definitions of Einstein radius may lead to some inconsistency, we find that the inferred Einstein radius $R_{\rm E,el}$ is a fairly good estimate of $R_{\rm E}$. The median bias (\ie, the median value of the biases for a large sample of systems) in $R_{\rm E}$ estimation is typically subpercent, regardless of lens mass models or noise levels, which, however, have a large effect on the scatter of the biases. The BPL-rigid model, which is found to be more accurate than the SIE, SPL, and BPL-free models, can determine the Einstein radius to an accuracy of $5\%$ or better, depending on the quality of extracted lensed images.

An accurate measurement of the Einstein radius implies an accurate measurement of the enclosed mass within it. This is indeed the case, as shown in Figure \ref{Fig:mkappa}. We find that all of the lens mass models investigated in this paper can measure the radial mean convergence around Einstein radius with controllable biases. However, away from Einstein radius, the mean convergence profiles are hard to constrain unless rigid priors are added to restrict the BPL model. We demonstrate that the BPL-rigid model can recover the mean convergence profiles quite well, with median biases less than $\sim5\%$ within $3~R_{\rm E}$ and less than $\sim10\%$ within $10~R_{\rm E}$.

As for the radial convergence profiles, we find that the median bias at Einstein radius for the SIE, SPL, and BPL-free models is typically larger than $10\%$, along with much larger scatters. For the BPL-rigid model, the median bias of the convergence at Einstein radius is typically less than $4\%$ with $\sim10\%$ scatters. It is difficult to constrain the radial convergence profiles over a broad range of radii based solely on lensed images. Among the lens mass models investigated, we find that the BPL-rigid model is the most successful one in recovering the radial convergence profiles, with median biases typically no more than $5\%$ and $10\%$ within $R_{\rm E}$ and $3R_{\rm E}$, respectively.

We also inspect the source light reconstructions in detail. We notice that, although the scatters on the biases are significant, there are basically no median biases in the estimation of centers, position angles, and axis ratios of the source galaxies. We find that the lens light contamination can significantly bias the estimation of the radial profile of source light distribution, \eg, the central light intensity $I_0$, effective radius $R_{\rm eff}$, and S\'ersic index $n$. The median biases in $I_0$ and $n$ can be largely reduced when the lens light contaminations are insignificant. However, the inference of $R_{\rm eff}$ is also sensitive to lens mass models in addition to the lens light contaminations. We realize that, to a large extent, the center offsets of source galaxies can be attributed to the existence of constant deflection angles in the central region of lenses, which are mainly caused by the complexity of lens mass distribution, \eg, deviations from smoothness and elliptical symmetry.

We look into the consistency between predicted and true AL-weighted LOSVDs. The results show that there are clearly positive biases in the estimation of AL-weighted LOSVDs, and these positive biases cannot be effectively reduced by improving the quality of lensed images. For the BPL-rigid model, the positive bias can be effectively eliminated by accounting for the projection effect, leading to an accuracy of $\sim6\%$ scatter in the prediction of AL-weighted LOSVDs. We thus conclude that, with the BPL-rigid model, it may be feasible to evaluate the external convergence to within $12\%$ accuracy for an SL system by comparing its true AL-weighted LOSVD with that predicted from the reconstructed lens mass distribution.

In short, we notice that a good fit to the lensed images does not necessarily indicate a good measurement of the lens mass or source light distribution. With suitable priors, the BPL model can significantly outperform the single power-law models in the reconstruction of lens mass and source light distributions, as well as the prediction of AL-weighted LOSVDs. In any case, we find that the Einstein radius cannot be constrained with a scatter of biases less than $2\%$ statistically by the smooth lens mass models investigated in this paper. Because of the large scatter of biases on the convergence measurement at Einstein radius, the fractional error in $H_0$ is unlikely to be much smaller than $10\%$ for a single SL time-delay system.

Finally, there are some issues to mention. One issue concerns the mock lenses that are picked out from Illustris galaxies. We know that the Illustris galaxies have large flat cores due to the softening effect, and are much larger than observed galaxies \citep{2017MNRAS.467.2879B,2017MNRAS.469.1824X}. The existence of flat cores largely increases the probability of forming central images and increases the size of the deviations of mass distribution from the SIE and SPL models in the central region. Thus, some relevant biases reported in this paper for the SIE and SPL models may be overestimated. On the other hand, if there are no actual central images, it may be more difficult to measure the lens mass distribution in the central region based solely on lensed images because of lensing degeneracies. The accuracy of lens mass measurement depends on how well the lens model is representative of the true lens mass distribution. Nonetheless, we demonstrate that the BPL-rigid model that relies on priors from stellar mass or lens light distribution can reliably recover the lens mass distribution over a wide range of radii and is less sensitive to the presence or absence of central images.


In addition, the existence of a large flat core tends to make the Einstein radius of the mock lens smaller and the effective radius of the lens light distribution larger. So the lens light contamination may be too significant in our mock SL images. As shown in Figure \ref{Fig:rein}, there are a few outliers whose Einstein radii are remarkably underestimated or overestimated. We find that most of them exhibit inconsistent lensing features with the true ones, demonstrating the significant effect of lens light contamination or the failure of the B-spline technique in extracting the lensed images of these cases. This also indicates that the criteria for defining ``grade-A'' lenses need to be improved. Nonetheless, as the actual SL sample size increases, there will be more lenses identified with smaller Einstein radii, which, as expected, will suffer more from lens light contamination. In a sense, we have addressed the significant effect of lens light contamination and demonstrated the success of the BPL-rigid model in dealing with SL systems with obvious lens light contamination.

There are some simplifications in generating and analyzing the mock images. For example, we do not take into consideration the lensing effects of environment and cosmological large-scale structures, which would complicate the data analyses and the lens mass model fittings. We neglect the redshift distributions of foreground lenses and background sources. When applying the B-spline technique to fit the lens light, feature masks are defined automatically by pixels with more than $5\%$ luminosity excess, which in practice cannot be recognized so easily. In the prediction of AL-weighted LOSVDs, the velocity anisotropy parameter is assumed to be constant and known for each lens. These simplifications or assumptions may, to some extent, have improved the measurement accuracies of the lens mass distribution or the AL-weighted LOSVDs.

Priors are essential for the BPL model to ensure the accuracy of lens mass reconstructions. We inspect the priors using the Illustris galaxies at a fixed redshift, while the priors are most likely redshift- and wavelength-dependent. There may exist more precise scaling relations between overall mass and stellar mass (or lens light) distributions. Thus, it will be worthwhile investigating the priors in more detail.

\section*{Acknowledgments}
We thank the referee for providing detailed comments and suggestions. W.D. acknowledges the support from the National Natural Science Foundation of China (NSFC) under grants 11803043, 11890691, and 11720101004. L.P.F acknowledges the support from NSFC grant 11933002, and  the Innovation Program 2019-01-07-00-02-E00032 of SMEC. Y.S. acknowledges the support from the China Manned Spaced (CMS) project No. CMS-CSST-2021-A12. R.L. acknowledges the support by National Key R\&D Program of China (No. 2022YFF0503403), the support of NSFC (Nos. 11988101 and 12022306), the support from the Ministry of Science and Technology of China (No. 2020SKA0110100), CAS Project for Young Scientists in Basic Research (No. YSBR-062), and the support from K.C.Wong Education Foundation. Z.H.F is supported by NSFC under grants 11933002 and U1931210. C.G.S. acknowledges the support by NSFC under grant No. 12141302, and the Science and Technology Commission of Shanghai Municipality (STCSM No. 22590780100). We acknowledge the science research grants from the CMS Project with Nos. CMS-CSST-2021-A01 and CMS-CSST-2021-B01. We would like to thank the Laohu high-performance computing (HPC) cluster supported by National Astronomical Observatories, Chinese Academy of Sciences, which was utilized for part of the MCMC runs. The authors also acknowledge Beijing PARATERA Tech CO., Ltd. (\url{https://www.paratera.com/}) for providing HPC resources that have contributed to the research results reported within this paper.


\end{document}